\documentclass[twocolumn]{revtex4-2}
\usepackage{graphicx,color}
\usepackage{xtab,afterpage}
\usepackage{lipsum}
\usepackage{bm}
\usepackage{amssymb,amsmath,amsfonts, mathrsfs}

\newcommand{\be}{\begin{equation}}
\newcommand{\ee}{\end{equation}}
\newcommand{\ba}{\begin{array}}
\newcommand{\ea}{\end{array}}
\newcommand{\bqa}{\begin{eqnarray}}
\newcommand{\eqa}{\end{eqnarray}}

\begin{document}



%
%

\title{Two-pole structures in a relativistic Friedrichs-Lee-QPC
scheme}


\author{Zhi-Yong Zhou}
\email[]{zhouzhy@seu.edu.cn}
\affiliation{School of Physics, Southeast University, Nanjing 211189,
P.~R.~China}
\author{Zhiguang Xiao\footnote{Corresponding author}}
\email[]{xiaozg@ustc.edu.cn}
\affiliation{NSFC-SFTP Peng Huanwu Center for Fundamental Theory and Interdisciplinary Center for Theoretical Study, University of Science
and Technology of China, Hefei, Anhui 230026, China}


\date{\today}

\begin{abstract}
A general appearance of two-pole structures is exhibited in a
relativistic Friedrichs-Lee model combined with a relativistic quark
pair creation model in a consistent manner. This kind of two-pole structure
could be found when a $q\bar q$ state couples to the open-flavor
continuum state in the $S$ partial wave. We found that many enigmatic
states, such as $f_0(500)/\sigma$, $K_0^*(700)/\kappa$, $a_0(980)$,
$f_0(980)$, $D_0^*(2300)$, $D_{s0}^*(2317)$, and $X(3872)$, together
with another higher state for each, all result from this kind of
two-pole structures. Furthermore, an interesting observation is that
this kind of two-pole structure will contribute roughly a total of
180$^\circ$ phase shift for the scattering process in a single channel
approximation.  This relativistic scheme may provide more insights
into the understanding of the properties of non-$q\bar q$ state. It is
also suggested that such two-pole structure could be a common
phenomenon which deserves studying both from theoretical and
experimental perspectives.

\end{abstract}


\maketitle

\section{Introduction}
As is well-known, states in general are related to the poles of the scattering
matrix. In hadron physics, there are always cases that two poles
appear together and dynamically related to each other in some
scattering processes.
For example, the $N(1405)$ signal in the $\bar KN$ and
$\pi \Sigma$ system was proposed to be
contributed by two poles dynamically generated on the same
Riemann sheet~\cite{Oller:2000fj,Jido:2003cb,Magas:2005vu,Hyodo:2007jq}. There
could also be the cases that the two poles are located on  different
Riemann sheets and represent the same state, one of them being  the
resonance pole and the other the shadow pole. A typical example is
the $N(1440)$ state which comes with both a second-sheet pole and
a third-sheet shadow pole\cite{Doring:2009yv,Arndt:2006bf}.  However, there is another case where one of the
two poles comes from a seed state and the other so called ``companion pole''~\cite{Boglione:2002vv} is
dynamically generated from the interaction between the same seed
state and some continuum state.  These double poles may be separated far
from each other and be regarded as different states.
An old example is the Unitary
Quark Model
proposed by T\"ornqvist~\cite{Tornqvist:1995kr} to explain the lowest
scalar nonet. This mechanism could be the cause of the
generation of the mysterious $X(3872)$ and a heavier $X(3940)$, that is, the $X(3872)$ could be regarded as being dynamically
generated by the interaction between the $\chi_{c1}(2P)$ state and
the continuum $D\bar
D^*$~\cite{Kalashnikova:2005ui,Ortega:2009hj,Takizawa:2012hy,Coito:2012vf,Sekihara:2014kya,
Zhou:2017dwj,Zhou:2017txt,Giacosa:2019zxw}.  Similar idea is
also used in the studies on the $\psi(3770)$
meson~\cite{Coito:2017ppc}.   In present paper, we will show that the
appearance of this
kind of two-pole structures could be a common phenomenon in hadron
spectrum ranging from the states with light quarks to those with heavy quarks,
and a simple dynamical origin of these two-pole structure is revealed
in a relativistic constituent quark picture.

The quark potential models are usually regarded as a criterion to
characterize the observed hadron states for its generally successful predictions~\cite{Godfrey:1985xj}.
However, a famous long-standing  puzzle is about the lightest $0^+$
scalar mesons which lie below 1.0 GeV, while the
 lowest $0^+$ scalar $(u\bar u\pm d\bar d)/\sqrt{2}$, $u\bar s$, and $s\bar s$
states predicted by the quark potential model are at about $1.1\sim 1.5$
GeV~\cite{Godfrey:1985xj}. The light scalar states observed in experiments are categorized in three groups
according to their isospins: (1) Five $I=0$ states: $f_0(500)$,
$f_0(980)$, $f_0(1370)$, $f_0(1500)$, and $f_0(1710)$; (2) Two $I=1/2$
states: $K_0^*(700)$ and $K_0^*(1430)$; (3) Two $I=1$ states:
$a_0(980)$ and $a_0(1450)$~\cite{Tanabashi:2018oca}. The attempt to categorize these scalar
states into suitable nonets was disturbed by the controversy about the
existence of $f_0(500)$ and $K_0^*(700)$, until their
poles were determined by model-independent
methods~\cite{Caprini:2005zr,Zhou:2004ms,Zheng:2003rw,DescotesGenon:2006uk,Pelaez:2020uiw,Pelaez:2015qba,Yao:2020bxx}.
Nowadays, $f_0(500)$,
$K_0^*(700)$, $a_0(980)$, and $f_0(980)$ are suggested to form
a $non$-$q\bar q$ nonet, while $f_0(1370)$, $K_0^*(1430)$,
$a_0(1450)$, and $f_0(1500)$(or $f_0(1710)$) are assumed to form the
$q\bar q$ nonets~\cite{Close:2002zu}. The lower nonet has an
``inverse" mass relation, which could be understood in the tetraquark
model proposed by Jaffe~\cite{Jaffe:1976ig,Maiani:2004vq}.
Another
method to study these states is to restore the resonance information
from the scattering amplitudes in the chiral perturbation
theory~($\chi$PT) with some unitarization schemes. Large $N_c$
analyses demonstrate that those states below 1.0 GeV really do not
behave like the $q\bar
q$ states~\cite{Oller:1998hw,Guo:2011pa,Pelaez:2003dy}.
Besides the puzzle of the light scalar states, more recently, some
hadron states with heavy quarks, such as $D_0^*(2300)$,
$D_{s0}^*(2317)$, and $X(3872)$, are also puzzling states which could
hardly be accommodated in the predicted $q\bar q$ states in the quark
potential model. Usually, by unitarizing the scattering amplitudes in
the heavy quark chiral perturbation theory~(H$\chi$PT), several groups
claimed that they could be regarded as the hadronic molecular
states\cite{Guo:2006fu,Guo:2006rp,Guo:2015dha,Albaladejo:2016lbb,Wu:2019vsy}.
Some other attempts were also made in pursuing the idea that they might be dynamically generated due to
the coupling of fundamental states to the continuum
states~\cite{vanBeveren:1982qb,vanBeveren:1983td,vanBeveren:1986ea,vanBeveren:2006ua,vanBeveren:2003kd,Coito:2012vf,Heikkila:1983wd,Tornqvist:1995kr,Zhou:2010ra,Boglione:1997aw,Boglione:2002vv,Geiger:1992va,Badalian:2020wua,Kalashnikova:2005ui,Ortega:2009hj,Takizawa:2012hy,Sekihara:2014kya,
Wolkanowski:2015jtc,Wolkanowski:2015lsa}.
The basic spirit of these models here is that by coupling a seed $q\bar q$
state to the
two-meson continuum states, another state could be
generated which could be identified as some possible $non$-$q\bar q$ states.
Along these lines,  we will  provide an
 understanding of all the above possible $non$-$q\bar q$ states
together using the  unified Friedrichs-Lee-QPC scheme.

In order to do so,
relativistic effects should be considered in a consistent way to cover
both the light and the heavy mesons.  Recently, we proposed the
relativistic Friedrichs-Lee-QPC scheme~\cite{Zhou:2020vnz} which combines
the exactly solvable relativistic Friedrichs-Lee
model~\cite{Antoniou:1998JMP} and the
relativistic quark pair creation~(RQPC) model~\cite{Fuda:2012xd} to
study both the low lying meson spectrum and the heavy meson
spectrum consistently.  In the present paper, this scheme will be
applied to study the hadron resonant states  mentioned above, in order
to understand their nature in this unified framework. We will show
that all of these states could be dynamically generated by coupling a
seed state and a continuum state, and together with the state
originated from the seed state for each one, they form two-pole
structures.

The paper is organized as follows: The relativistic Friedrichs-Lee-QPC
scheme is briefly reviewed in Section II. Section III is devoted to the
main numerical results and discussions on these hadron states. The
last section is the summary.

\section{A brief review of relativistic Friedrichs-Lee-QPC scheme}

The basic idea of the Friedrichs-Lee model is that, when the coupling
between a discrete state and a continuum state is considered, the
discrete state will dissolve into the continuum and becomes a
resonant state~\cite{Friedrichs:1948,Lee:1954iq}. There are
also other models implementing  similar idea in various areas in physics, such as in atomic
physics~\cite{PhysRevLett.86.2699} and in quantum
optics~\cite{JaynesCummings}. The Friedrichs-Lee model is also linked
to quantum field theory~(QFT) in~\cite{Giacosa:2011xa} and also used in the context of
baryons~\cite{Liu:2015ktc} and thermal systems~\cite{Lo:2019who}.

In fact, in the Friedrichs-Lee model, besides the pole shifted from
the original discrete state, other dynamically generated poles could
appear in the scattering amplitude~\cite{Xiao:2016dsx,Zhou:2017dwj}.
The characteristics of these dynamically generated states is that when
the coupling is being turned off, they will not go back to the bare
discrete state, but move to the singularities of the form factor,
which may be located at the infinity or in the complex plane, thus
justifying a dynamic nature of these states.  The form factor, as a
complex analytic function after analytic continuation with respect to
one complex variable, as long as it is not a constant, must have one
or more singularities somewhere on the complex plane, as the Liouville
theorem in complex analysis tells us. Thus, this kind of scenario of
the dynamically generated states could be a general mechanism.

This Friedrichs-Lee scheme was extended to a totally relativistic
scenario by including the relativistic kinematics and introducing the
creation and annihilation operators for a single-particle state and
for a two-particle continuum state mimicked by a so-called bilocal
field~\cite{Antoniou:1998JMP,Zhou:2020vnz}.  The full Hamiltonian of the
system is written down as
\bqa
P_0&=&\int \mathrm{d}^3\mathbf{p}\beta(E)\mathrm{d}E
EB^\dag(E,\mathbf{p})B(E,\mathbf{p})\nonumber\\
&+&\int
\mathrm{d}^3\mathbf{p}\omega(\mathbf{p})a^\dag(\mathbf{p})a(\mathbf{p})+\int
\mathrm{d}^3\mathbf{p}\beta(E)\mathrm{d}E \alpha(E)\nonumber\\
&&\times(a(\mathbf{p})+a^\dag(-\mathbf{p}))(B^\dag(E,\mathbf{p})
+B(E,-\mathbf{p})),
\label{eq:P0}
\eqa
where $\omega(\mathbf{p})$ denotes the energy of the single particle
 with momentum $\mathbf p$, $a^\dag(\mathbf{p})$ and
$a(\mathbf{p})$ being the operators to create or annihilate this single particle. $B^\dag(E,\mathbf{p})$ and
$B(E,\mathbf{p})$ are the so-called bilocal field creation operators  and annihilation
operators~\cite{Antoniou:1998JMP} introduced to mimic a
two-particle state creation and annihilation with total energy $E$ and
momentum $\mathbf p$. The
other internal quantum numbers such as  orbital angular momentum and
spins are suppressed.  $\alpha(E)$ is the coupling form factor
between
the single-particle state and the two-particle state, while
$\beta(E)\mathrm{d}E\mathrm{d}^3\mathbf p$ indicates the integration
measure of the two-particle state.

The eigenvalue problem is resolved by
finding the solution of operator $b^\dag(E,\mathbf{p})$ in
 \bqa
[P_0,b^\dag(E,\mathbf{p})]=p_0 b^\dag(E,\mathbf{p}),
\label{eq:eigeneq}
\eqa
and this equation could be exactly
solved by using the standard techniques of Bogoliubov
transformation~\cite{Zhou:2020vnz,Antoniou:1998JMP}.  With the exact
solution,
 the elastic scattering $S$-matrix 
of the two-particle continuum state can be expressed as
\begin{align}
S(E,\mathbf p;E',\mathbf p')=\delta^{(3)}(\mathbf p-\mathbf p')\delta(E-E')\Big(1-2\pi
i\frac{\rho(s)}{\eta_+(s)}\Big)\,,
\end{align}
in which $\eta_+(s)$, the inverse of resolvent function,
 reads
\bqa\label{etafunction}
&&\eta_+(s)=s-\omega_0^2-\Pi(s),\nonumber\\
&&\Pi(s)=\int_{s_{th}} \mathrm{d}s' \frac{\rho(s')}{s-s'+i0}.
\eqa
$s_{th}$ is the threshold value of $s$, the invariant mass squared of
the continuum state, and $\omega_0$ is the bare mass of the single
particle.  The spectral function $\rho(s)$ is defined as
$\rho(s)=2\omega_0\frac{k(s)E_1(s)E_2(s)}{\sqrt{s}}\alpha(s)^2$, in
which $k(s)$ is the magnitude of the relative momentum of the two particles in
their c.m. frame, and $E_1(s)$ and $E_2(s)$ are their respective
energies. As a consequence of introducing the annihilation operators,
Eq.(\ref{etafunction}) depends on $s$ instead of $E$, similar to the
relativistic dispersion relation, which is
different from its counterpart in the non-relativistic
case~\cite{Xiao:2016mon,Zhou:2017dwj}. This $\eta_+(s)$ function has a right hand
cut starting from the threshold, and could be analytically continued
to the complex $s$-plane with two Riemann sheets  and is then denoted
as $\eta(s)$. With the
imaginary part of the $\eta(s)$ function being just $\pi i\rho(s)$ above the
threshold, it is easy to see that the $S$ matrix is automatically
unitary. Since $\eta(s)$ is the
denominator of the $S$-matrix, the zero points of
the function on the Riemann sheets are just related to the virtual-state, bound-state, and
resonance-state poles of the scattering amplitude. The
left-hand-cut contribution is not included in this model, thus the crossing symmetry is
violated. However, since the left-hand cut only provides a smooth
background to the amplitude near the physical region, its effect can
be absorbed into the coupling constant and  would not
 have too much effect on
the pole positions which are our main interests in this paper.
In principle, the scheme could be extended to the cases with
multiple continua. Here we consider only the
elastic scattering cases with only one continuum state for simplicity.

The coupling form factor  $\alpha(s)$ describes how the bare discrete
one-particle state interacts with the bare two-particle continuum
state.  In our previous works~\cite{Zhou:2017dwj,Zhou:2017txt},
in the norelativistic Friedrichs model, the nonrelativistic QPC model
was used to describe such interactions of the mesons with heavy
quarks. The QPC model is related to the QCD since it just
parameterizes a kind of quark pair creation process from the vacuum
which can be viewed as a sub-process of QCD, which may be important in
the decay processes involving light
quarks~\cite{Dosch:1986dp,Kokoski:1985is,Ackleh:1996yt}. This model is
widely used in describing the interaction
of mesons in the literature and is tested to be reasonable (see for example~\cite{Micu:1968mk,LeYaouanc:1972vsx,Blundell:1995ev} and the
papers citing them). In present work, the RQPC model is adopted to include some
relativistic corrections to the QPC
model~\cite{Fuda:2012xd,Zhou:2020vnz}. Both at the
meson level and at the quark level, the relativistic canonical
one-particle and two-particle states and their transformation
properties under the Lorentz transformation between the c.m. frame and
the other frames are taken into
account~\cite{macfarlane1963,McKerrell1964,Fuda:2012xd}, so that the
whole scheme is formulated in a consistent relativistic manner.

In the RQPC model, to describe the bare meson coupling $A\rightarrow BC$, the $S$-matrix is defined as
\bqa
S_{fi}=I-2\pi i\delta(E_f-E_i)T,
\eqa
and then
\bqa
\langle BC|T|A\rangle=\delta^{(3)}(\mathbf{P}_A-\mathbf{P}_B-\mathbf{P}_C)\mathscr{M}^{m_{j_A}m_{j_B}m_{j_C}},
\label{eq:MABC}\eqa
where $|j_A,m_{j_A}\rangle$ are the quantum numbers of the total
angular momentum of $A$ and $\mathscr{M}^{m_{j_A}m_{j_B}m_{j_C}}$ is
the coupling amplitude.  Here, $|A\rangle$ corresponds to the bare
discrete state and $|BC\rangle $ corresponds to the bare continuum
state in the Friedrichs-Lee model.

We could write down a relativistic mock state of a meson, with
three-momentum $\mathbf{p}$, mass eigenvalue $\tilde W$,
orbital angular momentum $l$ of two quarks,  total spin $s$ of quarks,
 total angular momentum $j$ and its third component
$m_{j}$, as
\begin{align}\label{mockstate}
&|(\tilde W,^{2s+1}{l}_{{jm_{j}}})(\mathbf{p})\rangle=\sum_{m_lm_s}\sum_{m_1m_2}\sum_{m'_1 m'_2}\int \mathrm{d}^3\mathbf{k}\psi_{lm_{l}}(\mathbf{k})\nonumber\\
&\times\phi^{12}\omega^{12}|\mathbf p_1,s_1 m_1'\rangle\otimes|\mathbf p_2,s_2 m'_2\rangle\mathscr{D}^{s_1}_{m_1'm_1}[r_c(l_c(p),k_1)]\nonumber\\
&\times\mathscr{D}^{s_2}_{m_2'm_2}[r_c(l_c(p),k_2)]\langle s_1s_2m_1m_2|sm_{s}\rangle\langle lsm_{l}m_{s}|jm_{j}\rangle \nonumber\\
&\times N(\mathbf{p}_1,\mathbf{p}_2,\mathbf{p},\mathbf{k}),
\end{align}
where $\mathbf p_1$ and $\mathbf p_2$ denote the three momenta of
the quark and the antiquark in a general frame, $k_1$ and $k_2$ their four
momenta in the c.m. frame of the meson, $\phi^{12}$ the flavor wave
function, $\omega^{12}$ the color wave function of the meson,
and $\psi_{lm_{l}}(\mathbf{k})$ is the relative wave function of the quarks in
the c.m. frame. $\mathscr{D}^{s_1}_{m_1'm_1}[R]$ is the $SU(2)$
matrix representation of rotation group $R$, and $r_c(l_c(p),k_1)$
represents the Wigner rotation caused by two successive non-collinear
Lorentz boosts $l_c(k_1)$ and $l_c(p)$, which is the key difference between the relativistic meson mock state and the non-relativistic one~\cite{Blundell:1995ev}.
$N(\mathbf{p}_1,\mathbf{p}_2,\mathbf{p},\mathbf{k})$ is the
normalization factor.

The interaction operator for the quark pair creation could be expressed as~\cite{Fuda:2012xd,Micu:1968mk}
\begin{align}
&T=-\sqrt{8\pi}\gamma\int\frac{ \mathrm{d}^3\mathbf p_3\mathrm{d}^3\mathbf p_4}{\sqrt{\varepsilon_3(\mathbf p_3)\varepsilon_4(\mathbf p_4)}}\delta^3(\mathbf p_3+\mathbf p_4)\nonumber\\
&\times\sum_m\sum_{m_3m_4}\langle 1,m,1,-m|0,0\rangle\langle 1/2,m_3,1/2,m_4|1,-m\rangle \nonumber\\
&\times\mathscr{Y}_1^m(\frac{\mathbf p_3-\mathbf p_4}{2}) \phi_0^{34}\omega_0^{34}b^\dag_{m_3}(\mathbf p_3)d^\dag_{m_4}(\mathbf p_4),
\label{eq:T-Matrix}
\end{align}
where $\mathbf p_3$ and $\mathbf p_4$ denotes the three-momenta of the
quark and the anti-quark produced from the vacuum respectively,
$\phi_0^{34}$ and $\omega_0^{34}$ their flavor and color wave
functions, and $\gamma$ the quark pair production strength.
$\mathscr{Y}_l^m(\mathbf p)=|\mathbf p|^lY_l^m(\hat p)$ is the solid
harmonics. $b^\dag_{m_3}$ and $d^\dag_{m_4}$ are the creation
operators
of the quark and the anti-quark. Here, the spins and angular momentum
of quark and antiquark are combined to form a $J^{PC}=0^{++}$ quark
pair from the vacuum. The $\gamma$ parameter describes the
strength of generating the light quark pairs from the vacuum, which
was proposed to be universal in the original QPC model. However, many
factors could affect its value for different energy scales. First,
since the pair production process must be governed by QCD, the
strength would be affected by the running of the QCD coupling with the
energy scale. Second, the $\gamma$ parameter may also
effectively absorb
some interactions which are not included in the quark production
processes, such as the left-hand-cut effects. Thus, these effects may renormalize the $\gamma$ parameter
such that it may not be the same for low energy and high energy
processes. We first try a universal $\gamma$ parameter for both
the light and in the
heavy cases. We will see that for the system with heavier quarks, choosing another
$\gamma$ really causes the pole positions closer to the experimental results.  

 To connect to the large $N_c$ results, we note that by the large $N_c$
power counting, the three point amplitude of the mesons in
(\ref{eq:MABC}) should behave as $O(1/\sqrt{N_c})$. Since this
amplitude is proportional to $\gamma$, we can effectively put this
$N_c$ dependence into $\gamma$, i.e. $\gamma\propto1/\sqrt{N_c}$. Then
the large $N_c$ limit is the same as $\gamma\to 0$ limit. Similar
method is also used in \cite{Wolkanowski:2015jtc,Pelaez:2003dy}.

By the standard derivation from Eq.~(\ref{eq:MABC}),~(\ref{mockstate})
and~(\ref{eq:T-Matrix}), one could obtain the coupling amplitude
$\mathscr{M}^{m_{j_A}m_{j_B}m_{j_C}}$~\cite{Zhou:2020vnz}. Furthermore, if we choose the direction of meson $B$ along the $z$-direction, the
amplitude with the $BC$ system having relative angular momentum $L$ and the total spin
$S$ of meson pair is written down as~\cite{Jacob:1959at}
\begin{align}
\mathscr{M}^{LS}(\mathbf q_z)=&\frac{\sqrt{4\pi(2L+1)}}{2j_A+1}\nonumber\\
&\times\sum_{m_{j_B},m_{j_C}}\langle LS0(m_{j_B}+m_{j_C})|j_A(m_{j_B}+m_{j_C})\rangle\nonumber\\
&\times\langle j_B j_C m_{j_B}m_{j_C}|S(m_{j_B}+m_{j_C})\rangle\nonumber\\
&\times\mathscr{M}^{(m_{j_A}=m_{j_B}+m_{j_C})m_{j_B}m_{j_C}}(\mathbf q_z).
\label{eq:MLS}
\end{align}
Since this amplitude describes the interaction between the discrete
state and the continuum, we identify it as the coupling form factor $\alpha_{LS}(s)$ used in
 Eq.(\ref{eq:P0}) in the relativistic Friedrichs-Lee model in the center
of mass system.

The wave functions in the c.m. frame of the quark-antiquark system in
Eq.~(\ref{mockstate}) could be obtained by considering the
quark-antiquark interaction in the potential model, such as the
Godfrey-Isgur~(GI) model~\cite{Godfrey:1985xj} which is adopted here.
The relativistic effect is already implemented in the GI model such
that it is consistent with the relativistic Friedrichs-Lee scheme.
The wave functions for the bare states are then numerically solved
from GI's model. With these wave functions, the $\alpha(s)$ can be
obtained according to
Eq.~(\ref{eq:MLS}) by calculating the matrix element of $T$ in
Eq.~(\ref{eq:T-Matrix}) as in Eq.~(\ref{eq:MABC}). The interested reader are referred to our previous
work~\cite{Zhou:2020vnz} for details.
With all these preparations, the emergent physical states
could be obtained by finding out the zero points of the $\eta(s)$
function in Eq.~(\ref{etafunction}), which just relate to the poles of the scattering amplitude
of the continuum state. Furthermore, the physical phase shifts of the
scattering amplitude could also be represented as
$\delta(s)=\pi-\mathrm{Arg}[\eta(s)]$.
In principle, this relativistic scheme provide a consistent method
to study the meson state with light quarks or heavy quarks and  could
be viewed as including the hadron-loop corrections into the GI's
results.

\section{Numerical results and discussions}

The main purpose of this paper is to present the general emergence of
two-pole structures in meson spectrum
rather than making  a systematic fit. Thus, only one of the most
important continuum for every bare $q\bar q$ state is chosen.
Such a simplification to one-continuum scenario could also make the
two-pole picture  clearer  by avoiding the complexity introduced in
multi-continuum cases~\cite{Zhou:2011sp}.

{{
\begin{table*}\caption{\label{coupling} Correspondence of the discrete
states and the continuum states as the parameter
$\gamma=4.3$GeV. The values in the fourth column are the input mass of bare states. Unit is GeV.}
\begin{center}
\begin{tabular}{ccccccc}
\hline
\hline
``discrete"   & ``continuum"  & GI mass &  Input  & poles & experiment states & PDG values~\cite{Tanabashi:2018oca}\\
\hline
    $\frac{u\bar u+ d\bar d}{\sqrt{2}}(1^3P_0)$ & $(\pi\pi)_{I=0}$&
$1.09$& $1.3$ &  $\sqrt{s_{r1}}=1.34- 0.29 i$ & $f_0(1370)$ &
$1.35^{\pm0.15}- 0.2^{\pm 0.05}i$\\
      &   &  &   &  $\sqrt{s_{r2}}=0.39- 0.26 i$ & $f_0(500)$ &
$0.475^{\pm0.075}- 0.275^{\pm 0.075}i$ \\
\hline
   $u\bar s(1^3P_0)$  &  $(\pi K)_{I=\frac{1}{2}}$ & 1.23  & 1.42  &
$\sqrt{s_{r1}}=1.41- 0.17 i$ & $K_0^*(1430)$    &  $1.425^{\pm0.05}- 0.135^{\pm 0.04}i$  \\
    &     &   &     & $\sqrt{s_{r2}}=0.66- 0.34 i$  &  $K_0^*(700)$  &
$0.68^{\pm0.05}- 0.30^{\pm 0.04}i$   \\
\hline
   $s\bar s(1^3P_0)$  & $K\bar K$ & 1.35 & 1.68  &
$\sqrt{s_{r1}}=1.71- 0.16 i$  &  $f_0(1710)$   &  $1.704^{\pm0.012}- 0.062^{\pm 0.009}i$  \\
      &   &  &   &  $\sqrt{s_{b}}=0.98$,$\sqrt{s_{v}}=0.19$  &
$f_0(980)$   &  $0.99^{\pm0.02}- 0.028^{\pm 0.023}i$  \\
\hline
   $\frac{u\bar u- d\bar d}{\sqrt{2}}(1^3P_0)$  &  $\pi\eta$   & 1.09
&  1.3  &   $\sqrt{s_{r1}}=1.26- 0.14 i$    &  $a_0(1450)$   &
$1.474^{\pm0.019}- 0.133^{\pm 0.007}i$   \\
     &     &   &    & $\sqrt{s_{r2}}=0.70- 0.42 i$    &   $a_0(980)$
&  $0.98^{\pm0.02}- 0.038^{\pm 0.012}i$   \\
\hline
   $c\bar u(1^3P_0)$  &  $D\pi$   & 2.4   &  2.4  &
$\sqrt{s_{r1}}=2.58- 0.24 i$ &  $D_0^*(2300)$   &  $2.30^{\pm0.019}- 0.137^{\pm 0.02}i$   \\
     &     &   &    &   $\sqrt{s_{r2}}=2.08- 0.10 i$      &     &    \\
\hline
   $c\bar s(1^3P_0)$  &  $DK$   & 2.48   &  2.48  &
$\sqrt{s_{r1}}=2.80- 0.23 i$   &     &    \\
     &     &   &    &   $\sqrt{s_{b}}=2.24$,$\sqrt{s_{v}}=1.8$      &
$D_{s0}^*(2317)$   &  $2.317^{\pm0.0005}- 0.0038^{\pm 0.0038}i$   \\
\hline
   $b\bar u(1^3P_0)$  &  $\bar B \pi$   & 5.76   & 5.76  &
$\sqrt{s_{r1}}=6.01- 0.21 i$   &     &    \\
     &     &   &    &   $\sqrt{s_{r2}}=5.56- 0.07i$     &     &     \\
\hline
   $b\bar s(1^3P_0)$  &  $\bar B K$   & 5.83   &  5.83  &
$\sqrt{s_{r1}}=6.23- 0.17 i$   &     &   \\
     &     &   &    &   $\sqrt{s_{b}}=5.66$,$\sqrt{s_{v}}=5.3$      &     &     \\
\hline
   $c\bar c(2^3P_1)$  &  $D\bar D^*$   & 3.95   &  3.95 &
$\sqrt{s_{r1}}=4.01- 0.049 i$  &   $X(3940)$  &     \\
     &     &   &    &   $\sqrt{s_{b}}=3.785$    &  $X(3872)$   &   $3.87169^{\pm0.00017}$   \\
\hline
\hline
\end{tabular}
\end{center}
\end{table*}%
}}

Based on the theoretical background above,
the emergence of the two-pole structures in a rather broader spectrum could be studied.
The two-pole structure here means that, although there is only one bare
seed state interacting with the continuum state,
two sets of $S$-matrix poles related with each other appear. One set of poles
come from the bare seed state~(refers to as
``bare" poles) and the other set of poles are dynamically generated by
the interaction between the seed and the continuum state~(refers to
as ``dynamical"
poles). The
``dynamical'' poles in general may be originated from the singular point of the
interaction form factor, which may be located at infinity or faraway from the
seed~\cite{Xiao:2016dsx}.
 Usually, the ``bare" poles are a pair of complex conjugate
poles representing a resonance. The ``dynamical" poles could be a pair
of complex conjugate poles, a pair of bound-state and virtual-state
poles, or only one bound~(or virtual) state, depending on the coupling
form factor and its strength. %
 To demonstrate this  general
phenomenon, only one
Okuba-Zweig-Iizuka~(OZI) allowed continuum for each bare $q\bar q$
state is chosen.
Furthermore, the masses of some bare states are slightly deviated from
GI's prediction to
make the observables, such as the phase shifts, consistent with the
measured values in the experiments.
 This is reasonable, since GI's
calculations do not consider the interactions among the mesons. If
these interactions are included,
 and then do the same thing as GI, which is
similar to what we are doing here,
the masses should be ``renormalized'' from GI's results.
For our purpose, we will consider  nine  cases, ranging from the lowest
scalars to the one with one bottom quarks and the $c\bar
c(2^3P_1)$ cases related to $X(3872)$.

When the parameter $\gamma$ is chosen at about $4.3$GeV,  the poles of
the resolvents are extracted and listed in Table.~\ref{coupling}.
A general consistency between the pole positions, which are defined as
$\sqrt{s}=M- i\Gamma/2$, and the  states in
the PDG table could be found.
\begin{figure}
  \centering
  \includegraphics[width=5cm]{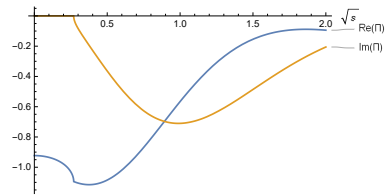}\\
  \caption{Real and imaginary parts of $\Pi(s)$ for the scalar $(u\bar u+d\bar d)/\sqrt{2}$ state.}\label{Pi}
\end{figure}

\begin{figure}
  \centering
  \includegraphics[width=8cm]{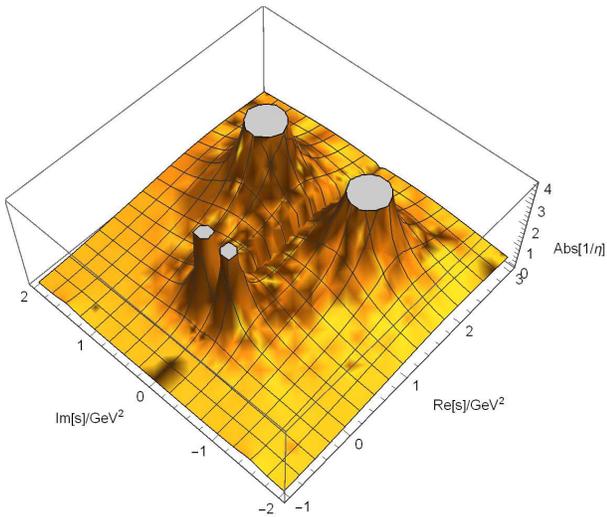}\\
  \caption{ $|1/\eta(s)|$ for the scalar $(u\bar{u}+d\bar{d})/\sqrt{2}$ on  the complex $s$ plane of the second Riemann sheet with $\gamma$=4.3 GeV.}\label{pole00}
\end{figure}

For the scalar $(u\bar u+d\bar d)/\sqrt{2}$ sector, the real
part and imaginary part of self-energy function of $\Pi(s)$  are
depicted in Fig.\ref{Pi}. Since there is an exponentially decaying factor in the
GI's wave function, the interaction form factor approaches zero fast enough in the
high energy limit such that the integration in the self-energy function
converges automatically. This is a general feature in our scheme.
The resolvent $1/\eta(s)$ has two pairs of poles on the second Riemann sheet when it is analytically continued to the complex-$s$ plane, as shown in Fig.\ref{pole00}.
The enigmatic broad $f_0(500)$  appears
naturally as the ``dynamical" state  represented by the lighter pair of poles in Fig.\ref{pole00}, while the ``bare" one is shifted to become
$f_0(1370)$ represented by the higher ones in Fig.\ref{pole00}. Similarly, $K_0^*(700)$ and $K_0^*(1430)$ are the
``dynamical" pole and the ``bare" pole of scalar $u\bar s$ states respectively.
The coupling of scalar $s\bar s$ bare state and the $K\bar K$
continuum, which is OZI-allowed, leads to a ``dynamical''
bound-state pole just below the $K\bar K$ threshold which corresponds
to the $f_0(980)$ and another virtual-state
pole at $0.19$ GeV which is too low to be observed.
If the $\pi\pi$ continuum coupling to $s\bar s$ is
considered, which is OZI-suppressed, there will be a new cut from the $\pi\pi$ threshold and
the bound state will move onto the second sheet and becomes a narrow resonance pole.
This general argument is in agreement with the characteristics
of $f_0(980)$, which appears as a peak structure in
$J/\psi\rightarrow\phi\pi^+\pi^-$ while being nearly invisible in
$J/\psi\rightarrow\omega\pi^+\pi^-$~\cite{Deandrea:2000yc}.
At the
same time, the bare $s\bar s$ is shifted onto the complex plane at
 $1.71- 0.16i$ GeV, whose properties is consistent with the
$f_0(1710)$, which has a large $s\bar s$ components as observed in
experiments~\cite{Acciarri:2000ex}.  To make $f_0(980)$ appear in the
$\pi\pi$ scattering channel, one must also consider the coupling of
the scalar $(u\bar u+d\bar d)/\sqrt 2$ to both the $\pi\pi$ and the
$K\bar K$ continua, and then the $f_0(980)$ will come into the
$\pi\pi$ interaction through $K\bar K$ loop, i.e. through the off-diagonal matrix elements of the inverse resolvent. This will not
affect the existence of the $f_0(500)$ and $f_0(1370)$ poles, but only
shift them a little. Since we are confining ourselves to the single
channel cases in our present paper, we will leave this for future work.

For the isovector scalar $(u\bar u-d\bar d)/{\sqrt{2}}$ state,
only coupling to the $\pi\eta$ continuum is considered in this work
and a broad resonance at about 0.7GeV is found. Actually, coupling to
$K\bar K$ continuum could be comparable with the coupling to $\pi\eta$, and
the $K\bar K$ threshold could truncate  the contribution of a broad
resonance pole to produce a narrow peak below the threshold as
illustrated in the Flatt\'e
effect~\cite{Flatte:1976xu,Zhou:2010ra,Tornqvist:1995kr}. A more
reasonable description might need a two-channel scenario, which is
beyond the scope of this work.

For the scalar $c\bar u$ and $c\bar s$ states, two-pole structures are
also found. Clearly, due to  coupling between the scalar $c\bar s$ state and the $DK$
continuum, the ``dynamical'' bound-state pole at 2.24 GeV could correspond to the $D_{s0}^*(2317)$ state.
The ``bare" pole originated
from the $c\bar s(1^3P_0)$ seed state is located at about $2.80- 0.23i$ GeV. The scalar $c\bar u$ state could also produce two poles
at $2.08- 0.10i$ GeV and $2.58- 0.24i$ GeV, in which the lower one
is the ``dynamical" one and the higher is the ``bare" one. Although the
experiment only claimed a broad resonance called $D_0^*(2300)$, it
could be contributed by two poles.  Further efforts to distinguish these two close poles are quite valuable,
because this occasion is different from other two-pole structures
whose poles separate from each other. Such a two-pole structure has
also been found by the calculations based on unitarizing the H$\chi$PT
amplitudes~\cite{Guo:2015dha,Albaladejo:2016lbb,Meissner:2020khl}, and they are also comparable with the LQCD simulation~\cite{Moir:2016srx}.
Similarly, their counterparts for the scalar $b\bar u$ and $b\bar s$
states could also be found here, and the two poles for each case are listed in Table.~\ref{coupling}.
 In the unitarized H$\chi$PT approach, the two poles are found
to come from two different $SU(3)$ multiplets in the $SU(3)$ limit. However, after the
$SU(3)$ breaking effect is turned
on, there should be mixing between these two kinds of $SU(3)$
multiplets and one can no longer distinguish them by different
representations. In our approach, we are working with broken $SU(3)$
from the beginning with different $s$ quark and $u/d$ quark masses and
the seed should also couple to both triplet and sextet of the
continuum.  Whether the two-pole structure here is really related to
the one in unitarized H$\chi$PT is not trivial and is beyond the scope of our present
work.

The $c\bar c(2^3P_1)$ state could also produce a two-pole structure as
we have shown in the non-relativistic Friedrichs
model~\cite{Zhou:2017dwj,Zhou:2017txt}. Here,  the ``dynamical" pole is also a bound
state below the $D\bar D^*$ threshold but much lower at about 3.78
GeV. 

The lower results of $X(3872)$ and $D_{s0}^*(2317)$ compared with
the observed values might indicate
that the $\gamma$ parameter for heavy
mesons may be different from the light states   as discussed before.  If the $\gamma$ parameter is chosen at $3.0$GeV to produce
the accurate mass of $D_{s0}^*(2317)$, the masses of the other states
are listed in Table.~\ref{tunedvalues}, which are worth pursuing in future
experiments.
The results of $c\bar c(2^3P_1)$ are also improved, closer to our previous
works~\cite{Zhou:2017dwj,Zhou:2017txt}.  Nevertheless, this still
demonstrates the existence of the two-pole structure.

\begin{table}\caption{\label{tunedvalues} The poles' positions for the heavy mesons as $\gamma=3.0$ GeV. Unit is GeV.}
\begin{center}
\begin{tabular}{ccc}
\hline
\hline
``states"   & bare poles  &   dynamical poles\\
\hline
   $c\bar u(1^3P_0)$  &   $\sqrt{s_{r1}}=2.39- 0.18 i$  &
$\sqrt{s_{r2}}=2.21- 0.28 i$        \\
\hline
   $c\bar s(1^3P_0)$  &  $\sqrt{s_{r1}}=2.68- 0.26 i$   &     $\sqrt{s_{b}}=2.32$,$\sqrt{s_{v}}=1.9$    \\
\hline
   $b\bar u(1^3P_0)$  &   $\sqrt{s_{r1}}=5.85- 0.26 i$   &
$\sqrt{s_{r2}}=5.62- 0.13 i$    \\
\hline
   $b\bar s(1^3P_0)$  &   $\sqrt{s_{r1}}=6.11- 0.22i$   &
$\sqrt{s_{b}}=5.72$,$\sqrt{s_{v}}=5.4$       \\
\hline
   $c\bar c(2^3P_1)$  & $\sqrt{s_{r}}=3.99-0.05 i$
   &   $\sqrt{s_{b}}=3.84$        \\
\hline
\hline
\end{tabular}
\end{center}
\end{table}%

\begin{figure}
  \centering
  \includegraphics[width=4cm]{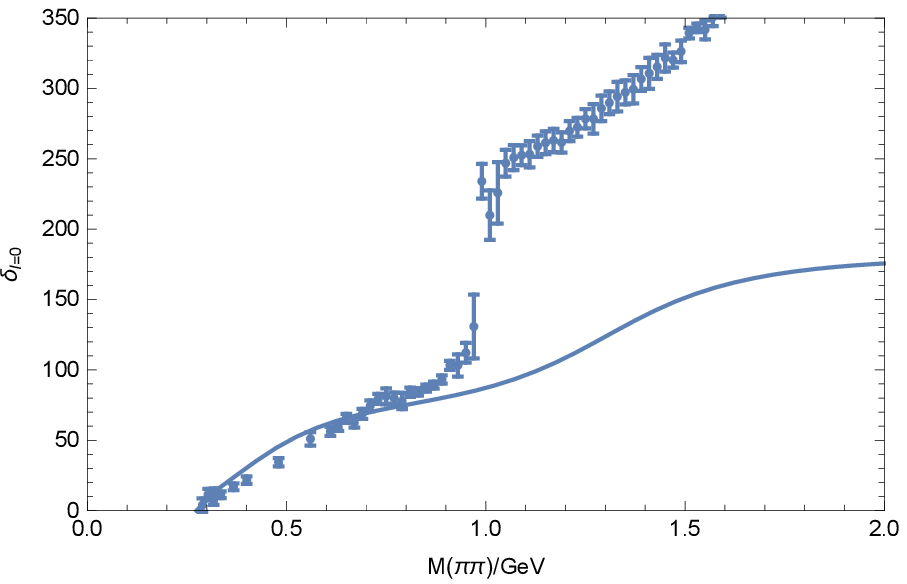}
  \includegraphics[width=4cm]{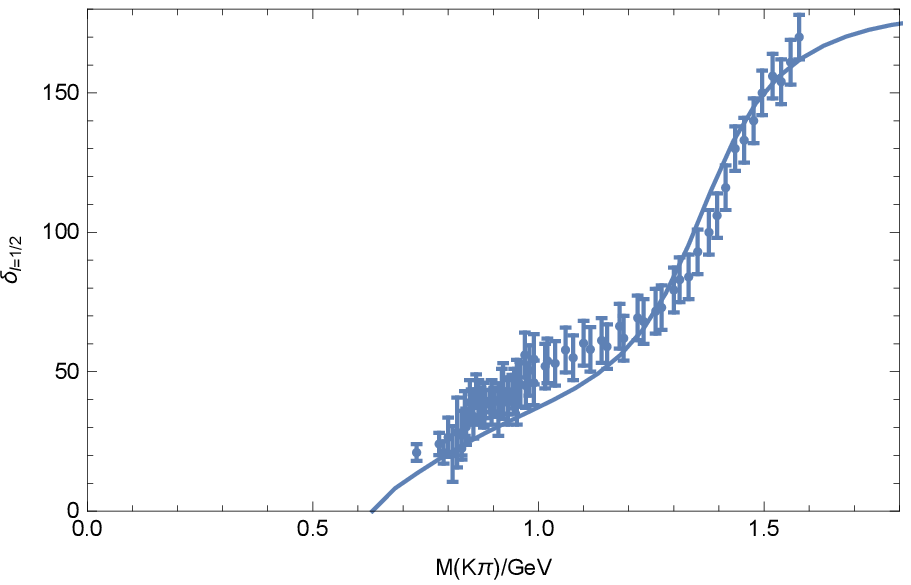}\\
  \caption{Comparisons of experimental phase shifts and the
theoretical calculations when $\gamma=4.3$GeV. The left one is that of $IJ=00$ $\pi\pi$ scattering and the right one is that of $IJ=\frac{1}{2}0$ $\pi K$ scattering. The solid line is the contribution of two-pole structure.}\label{phase}
\end{figure}

Besides the coincidence of the poles' positions with the experiment,
further evidences of two-pole structures come from the properties of
the scattering phase shifts. A careful analysis of this scheme shows a
sum rule for the single channel phase shift here,
$\delta(\infty)-\delta(s_{th})=180^\circ$~\cite{Zhou:2020vnz}. This means that  the two-pole
structure contributes a total phase shift of $180^\circ$, which should
generally not be satisfied if the two poles are
independent~\footnote{A frequently used parameterization of an
$S$-wave
resonance contribution to the $S$-matrix is $S=\frac
{s-M^2-i\rho(s)G_1
 }{s-M^2+i\rho(s)G_1
}$ which contributes $180^\circ$ to the phase shift, where
$\rho(s)=2k/\sqrt s$ is the kinematic factor, $G_1$ being a constant. However,
this representation only works for narrow resonances, since besides a
pair of resonance poles on the second Riemann sheet, there is
also a
virtual state pole. A more suitable parametrization is $S=\frac
{s-M^2-is\,\rho(s)G_2
}{s-M^2+is\,\rho(s)G_2
}$ which contributes less than $180^\circ$ (see \cite{Zheng:2003cr} for
details).}.
This property could be easily verified in $I=1/2$ $\pi K$ scattering
phase shift as shown in Fig.~\ref{phase}, where the $K_0^*(700)$
and $K_0^*(1430)$ contribute a total of roughly $180^\circ$ at
about 1.7GeV, consistent with
the extracted data from the $K\pi$ scattering~\cite{Mercer:1971kn,Estabrooks:1977xe,Bingham:1972vy,Aston:1987ir}.
The phase shift data of the $IJ=00$ $\pi\pi$ scattering also provide
some hints
to this sum rule, even though it is more complicated for being contributed
by both the states generated from  $(u\bar u+d\bar d)/\sqrt{2}$ and $s\bar s$ states.
As shown in Fig.~\ref{phase}, if the sharp rise of about $180^\circ$ at
about 1.0 GeV contributed by the narrow  $f_0(980)$ is removed, the experimental phase shift is just rising smoothly with a
total phase shift about $180^\circ$ at around 1.6GeV, which may
suggest  the same sum rule for $f_0(500)$ and $f_0(1370)$. It is
worth noting that the
unitarized $\chi$PT also suggests that only $\sigma$ or $\kappa$ would
not provide a phase shift of 180${}^\circ$~\cite{Dobado:1996ps}.

 Some further remarks about the sensitivity of the parameters is
in order. Notice that the wave functions of the bare meson states are
completely fixed by the GI model whose parameters are totally
determined by fitting the meson mass spectrum as in
ref.~\cite{Godfrey:1985xj}. Thus, in principle, there is only one free
parameter $\gamma$ that denotes the production strength of the $q\bar
q$ from the vacuum. Once $\gamma$ is chosen, the hadron resonance
poles will be determined at the same time. When $\gamma$ changes, all
the poles will change simultaneously. As we have mentioned above, to
make all the poles' position reasonable we chose $\gamma$ to be at 4.3
or 3.0 GeV.  The consistency with the experiment values can be seen from
Table.~\ref{coupling} and \ref{tunedvalues}, which is enough to
demonstrate our proposal that the two-pole structure could be a
general phenomenon.

\begin{figure}
  \centering
  \includegraphics[width=7cm]{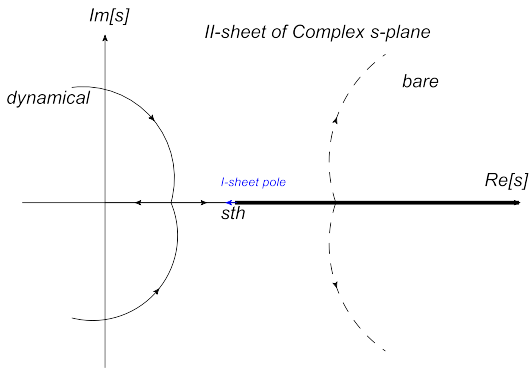}
  \caption{The general trajectories of two pairs of poles for the two-pole structures on the second Riemann sheet of the $s$-plane as $\gamma$ increases. When $\gamma$ becomes large enough, one of the dynamical pole will move across the threshold and become a bound-state pole as shown in a blue arrow line.}\label{twopole}
\end{figure}

It seems that the relative positions of the two poles in different cases
are different.
However, a numerical experiment could show some common behavior of
the pole trajectories as shown in Fig.~\ref{twopole}, which shed more light on the general properties
of the two-pole structure.
For all the  $J^{P}=0^{+}$ cases, as the coupling constant $\gamma$ increases from zero,
the ``bare" pole will move away from the real axis
to the second Riemann sheet and becomes a pair of conjugate resonance poles. At the same time,
another pair of ``dynamical" conjugate poles
come from the deep complex $s$-plane, and move towards the real
axis and meet on the axis below the threshold becoming a pair of
virtual-state poles when the coupling is large enough.
As the coupling
strength keeps increasing, one of the virtual state pole will move
down along the real axis and the other one moves up
across the threshold to
the first Riemann sheet and becomes a bound state.
On which part of the trajectory the pole
position is located will depend on the renormalization effect on $\gamma$, wave functions and the specific parameters
such as quark masses in each cases. For the scalar $s\bar s$,
$c\bar s$, and $b\bar s$ states, the coupling is so strong that a bound state and a virtual state are produced for
each case, but for other scalar cases the ``dynamical" poles remains to
be resonances.

Based on the above observation, we would extract some general features
in these two-pole phenomena. First, the appearance of a ``dynamical'' pole needs a nontrivial
form factor, which is produced by the coupling of mesons
composed of more fundamental quarks. Thus we reach our first
general statement:
 coupling the seed and continuum where all particles  involved are composite particles
would always produce ``dynamical'' poles such that this kind of
two-pole structure is possible.
We have also seen another feature: as the coupling increases from zero,
the ``dynamical'' resonance poles come from
far away towards the physical region and may become bound states or
nearby virtual states when the coupling becomes stronger. The
``dynamical''
pole and the ``bare'' pole may contribute a total phase shift of about $180^\circ$ if the single
channel approximation can be applied.

Other approaches in the literature also support
such a phenomenon. In $\chi$PT, one can also introduce intermediate
$s$-channel resonance in the meson scattering, and after unitarization there
could also be dynamically generated accompanying states, such as in
\cite{Oller:1998zr}, where $f_0(600)/\sigma$, $\kappa$, $a_0(980)$ and
$f_0(980)$ are dynamically generated by coupling a nonet around
1.0GeV and 1.4GeV to the low
lying pseudoscalars. Another effective field theory
method used by Wolkanowski et.al.\cite{Wolkanowski:2015lsa,
Wolkanowski:2015jtc, Giacosa:2019ldb, Giacosa:2019zxw} also presented such a phenomenon,
where the name ``companion pole'' is used.
Their results are similar to ours here. In particular, the
pole trajectories and the phase shifts  for the $K^*_0(700)$ and
$K_0^*(1430)$ in
ref.~\cite{Wolkanowski:2015jtc} also demonstrate the
similar general features listed in the previous paragraph.
In quark model, Beveran and Rupp's
approach~\cite{vanBeveren:1982qb,vanBeveren:1983td,vanBeveren:1986ea,vanBeveren:2006ua,vanBeveren:2003kd}
is in the same spirit as ours. Instead of using the Friedrichs-Lee model they
directly solved the Schr\"odinger equation with the potential extracted
from the QPC model or put by hand, and
their relativistic extension of the QPC model~\cite{vanBeveren:1982qb,vanBeveren:1983td} is different from ours.
However, their results and pole trajectories are also similar to ours
here and thus demonstrates the general features listed above. More
recent works~\cite{Lukashov:2019dir,Badalian:2020wua} using the so
called chiral confining Lagrangian  also produce similar results for
the lightest scalars. 

\section{Summary}

To sum up, the interaction between discrete states and the continuum
in general may dynamically generate new states and thus results in the
two-pole structures, which may be a
general mechanism in the strong interactions among hadrons. We show
here that the light $0^+$ scalars $f_0(500)/f_0(1370)$, $f_0(980)/f_0(1710)$, $K^*_0(700)/K^*_0(1430)$, $a_0(980)/a_0(1450)$,
are this kind of two-pole structures by using the relativistic
Friedrichs-Lee-QPC scheme. Furthermore, the two-pole structures
 $D_0^*(2210)/D_0^*(2390)$, $D_{s0}^*(2317)/D_{s0}^*(2680)$ and their counterparts with $b$ quark
are also suggested for the
future experiments to explore.
Though the single channel assumption and the limitation of RQPC may
bring in some uncertainty in the pole positions, our results for a
wide range of resonances being consistent with the experiment are enough to show
the general phenomenon of two-pole structure.  Some of the similar two-pole
structures were also found in the unitarized H$\chi$PT
approach~\cite{Guo:2015dha,Albaladejo:2016lbb,Meissner:2020khl}.
However, our
mechanism here provides a unified constituent quark picture on how
these two-pole structures is generated. It is surprising that these
two very different approaches converge to the similar results.
These are two very different pictures and it is hard to tell
which picture is correct or wrong and to make a simple comparison. Or
maybe both pictures are partially correct and are just dual to each
other in describing of the same physics from different facets.
Besides the two-pole structures listed above, there could be other
two-pole structures generated by the same mechanism which should be
paid attention to and be searched for in future experiments.
Furthermore, this mechanism could be much more general beyond hadron
physics. In molecular physics, atomic physics or condense matter
physics, there could also be composite particles scattering where a
seed couples with two-particle continuum, and such two-pole structure
may also be present.  Further exploration in this direction in these
areas may reveal new phenomena in future.

\textit{Acknowledgement:} Helpful discussions with  Hai-Qing Zhou, Zhi-Hui Guo, Gang Li, and Feng-kun Guo are appreciated. This work is supported by China National Natural
Science Foundation under contract No. 11975075, No. 11575177, and No.11947301. Z.Z is also supported by the Natural Science Foundation of Jiangsu Province of China under contract No. BK20171349.

\bibliographystyle{apsrev4-1}
\bibliography{Ref}

\begin{thebibliography}{86}%
\makeatletter
\providecommand \@ifxundefined [1]{%
 \@ifx{#1\undefined}
}%
\providecommand \@ifnum [1]{%
 \ifnum #1\expandafter \@firstoftwo
 \else \expandafter \@secondoftwo
 \fi
}%
\providecommand \@ifx [1]{%
 \ifx #1\expandafter \@firstoftwo
 \else \expandafter \@secondoftwo
 \fi
}%
\providecommand \natexlab [1]{#1}%
\providecommand \enquote  [1]{``#1''}%
\providecommand \bibnamefont  [1]{#1}%
\providecommand \bibfnamefont [1]{#1}%
\providecommand \citenamefont [1]{#1}%
\providecommand \href@noop [0]{\@secondoftwo}%
\providecommand \href [0]{\begingroup \@sanitize@url \@href}%
\providecommand \@href[1]{\@@startlink{#1}\@@href}%
\providecommand \@@href[1]{\endgroup#1\@@endlink}%
\providecommand \@sanitize@url [0]{\catcode `\\12\catcode `\$12\catcode
  `\&12\catcode `\#12\catcode `\^12\catcode `\_12\catcode `\%12\relax}%
\providecommand \@@startlink[1]{}%
\providecommand \@@endlink[0]{}%
\providecommand \url  [0]{\begingroup\@sanitize@url \@url }%
\providecommand \@url [1]{\endgroup\@href {#1}{\urlprefix }}%
\providecommand \urlprefix  [0]{URL }%
\providecommand \Eprint [0]{\href }%
\@ifxundefined \urlstyle {%
  \providecommand \doi  [0]{\begingroup \@sanitize@url \@doi}%
  \providecommand \@doi [1]{\endgroup \@@startlink {\doibase
  #1}doi:\discretionary {}{}{}#1\@@endlink }%
}{%
  \providecommand \doi  [0]{doi:\discretionary{}{}{}\begingroup
  \urlstyle{rm}\Url }%
}%
\providecommand \doibase [0]{http://dx.doi.org/}%
\providecommand \Doi [0]{\begingroup \@sanitize@url \@Doi }%
\providecommand \@Doi  [1]{\endgroup\@@startlink{\doibase#1}\@@Doi}%
\providecommand \@@Doi [1]{#1\@@endlink}%
\providecommand \selectlanguage [0]{\@gobble}%
\providecommand \bibinfo  [0]{\@secondoftwo}%
\providecommand \bibfield  [0]{\@secondoftwo}%
\providecommand \translation [1]{[#1]}%
\providecommand \BibitemOpen [0]{}%
\providecommand \bibitemStop [0]{}%
\providecommand \bibitemNoStop [0]{.\EOS\space}%
\providecommand \EOS [0]{\spacefactor3000\relax}%
\providecommand \BibitemShut  [1]{\csname bibitem#1\endcsname}%
\bibitem [{\citenamefont {Oller}\ and\ \citenamefont
  {Mei{\ss}ner}(2001)}]{Oller:2000fj}%
  \BibitemOpen
  \bibfield  {author} {\bibinfo {author} {\bibfnamefont {J.}~\bibnamefont
  {Oller}}\ and\ \bibinfo {author} {\bibfnamefont {U.~G.}\ \bibnamefont
  {Mei{\ss}ner}},\ }\Doi {10.1016/S0370-2693(01)00078-8} {\bibfield  {journal}
  {\bibinfo  {journal} {Phys. Lett. B},\ }\textbf {\bibinfo {volume} {500}},\
  \bibinfo {pages} {263} (\bibinfo {year} {2001})},\ \Eprint
  {http://arxiv.org/abs/hep-ph/0011146} {arXiv:hep-ph/0011146} \BibitemShut
  {NoStop}%
\bibitem [{\citenamefont {Jido}\ \emph {et~al.}(2003)\citenamefont {Jido},
  \citenamefont {Oller}, \citenamefont {Oset}, \citenamefont {Ramos},\ and\
  \citenamefont {Mei{\ss}ner}}]{Jido:2003cb}%
  \BibitemOpen
  \bibfield  {author} {\bibinfo {author} {\bibfnamefont {D.}~\bibnamefont
  {Jido}}, \bibinfo {author} {\bibfnamefont {J.}~\bibnamefont {Oller}},
  \bibinfo {author} {\bibfnamefont {E.}~\bibnamefont {Oset}}, \bibinfo {author}
  {\bibfnamefont {A.}~\bibnamefont {Ramos}}, \ and\ \bibinfo {author}
  {\bibfnamefont {U.}~\bibnamefont {Mei{\ss}ner}},\ }\Doi
  {10.1016/S0375-9474(03)01598-7} {\bibfield  {journal} {\bibinfo  {journal}
  {Nucl. Phys. A},\ }\textbf {\bibinfo {volume} {725}},\ \bibinfo {pages} {181}
  (\bibinfo {year} {2003})},\ \Eprint {http://arxiv.org/abs/nucl-th/0303062}
  {arXiv:nucl-th/0303062} \BibitemShut {NoStop}%
\bibitem [{\citenamefont {Magas}\ \emph {et~al.}(2005)\citenamefont {Magas},
  \citenamefont {Oset},\ and\ \citenamefont {Ramos}}]{Magas:2005vu}%
  \BibitemOpen
  \bibfield  {author} {\bibinfo {author} {\bibfnamefont {V.}~\bibnamefont
  {Magas}}, \bibinfo {author} {\bibfnamefont {E.}~\bibnamefont {Oset}}, \ and\
  \bibinfo {author} {\bibfnamefont {A.}~\bibnamefont {Ramos}},\ }\Doi
  {10.1103/PhysRevLett.95.052301} {\bibfield  {journal} {\bibinfo  {journal}
  {Phys. Rev. Lett.},\ }\textbf {\bibinfo {volume} {95}},\ \bibinfo {pages}
  {052301} (\bibinfo {year} {2005})},\ \Eprint
  {http://arxiv.org/abs/hep-ph/0503043} {arXiv:hep-ph/0503043} \BibitemShut
  {NoStop}%
\bibitem [{\citenamefont {Hyodo}\ and\ \citenamefont
  {Weise}(2008)}]{Hyodo:2007jq}%
  \BibitemOpen
  \bibfield  {author} {\bibinfo {author} {\bibfnamefont {T.}~\bibnamefont
  {Hyodo}}\ and\ \bibinfo {author} {\bibfnamefont {W.}~\bibnamefont {Weise}},\
  }\Doi {10.1103/PhysRevC.77.035204} {\bibfield  {journal} {\bibinfo  {journal}
  {Phys. Rev. C},\ }\textbf {\bibinfo {volume} {77}},\ \bibinfo {pages}
  {035204} (\bibinfo {year} {2008})},\ \Eprint {http://arxiv.org/abs/0712.1613}
  {arXiv:0712.1613 [nucl-th]} \BibitemShut {NoStop}%
\bibitem [{\citenamefont {Doring}\ \emph {et~al.}(2009)\citenamefont {Doring},
  \citenamefont {Hanhart}, \citenamefont {Huang}, \citenamefont {Krewald},\
  and\ \citenamefont {Mei{\ss}ner}}]{Doring:2009yv}%
  \BibitemOpen
  \bibfield  {author} {\bibinfo {author} {\bibfnamefont {M.}~\bibnamefont
  {Doring}}, \bibinfo {author} {\bibfnamefont {C.}~\bibnamefont {Hanhart}},
  \bibinfo {author} {\bibfnamefont {F.}~\bibnamefont {Huang}}, \bibinfo
  {author} {\bibfnamefont {S.}~\bibnamefont {Krewald}}, \ and\ \bibinfo
  {author} {\bibfnamefont {U.-G.}\ \bibnamefont {Mei{\ss}ner}},\ }\Doi
  {10.1016/j.nuclphysa.2009.08.010} {\bibfield  {journal} {\bibinfo  {journal}
  {Nucl. Phys. A},\ }\textbf {\bibinfo {volume} {829}},\ \bibinfo {pages} {170}
  (\bibinfo {year} {2009})},\ \Eprint {http://arxiv.org/abs/0903.4337}
  {arXiv:0903.4337 [nucl-th]} \BibitemShut {NoStop}%
\bibitem [{\citenamefont {Arndt}\ \emph {et~al.}(2006)\citenamefont {Arndt},
  \citenamefont {Briscoe}, \citenamefont {Strakovsky},\ and\ \citenamefont
  {Workman}}]{Arndt:2006bf}%
  \BibitemOpen
  \bibfield  {author} {\bibinfo {author} {\bibfnamefont {R.}~\bibnamefont
  {Arndt}}, \bibinfo {author} {\bibfnamefont {W.}~\bibnamefont {Briscoe}},
  \bibinfo {author} {\bibfnamefont {I.}~\bibnamefont {Strakovsky}}, \ and\
  \bibinfo {author} {\bibfnamefont {R.}~\bibnamefont {Workman}},\ }\Doi
  {10.1103/PhysRevC.74.045205} {\bibfield  {journal} {\bibinfo  {journal}
  {Phys. Rev. C},\ }\textbf {\bibinfo {volume} {74}},\ \bibinfo {pages}
  {045205} (\bibinfo {year} {2006})},\ \Eprint
  {http://arxiv.org/abs/nucl-th/0605082} {arXiv:nucl-th/0605082} \BibitemShut
  {NoStop}%
\bibitem [{\citenamefont {Boglione}\ and\ \citenamefont
  {Pennington}(2002)}]{Boglione:2002vv}%
  \BibitemOpen
  \bibfield  {author} {\bibinfo {author} {\bibfnamefont {M.}~\bibnamefont
  {Boglione}}\ and\ \bibinfo {author} {\bibfnamefont {M.}~\bibnamefont
  {Pennington}},\ }\Doi {10.1103/PhysRevD.65.114010} {\bibfield  {journal}
  {\bibinfo  {journal} {Phys. Rev. D},\ }\textbf {\bibinfo {volume} {65}},\
  \bibinfo {pages} {114010} (\bibinfo {year} {2002})},\ \Eprint
  {http://arxiv.org/abs/hep-ph/0203149} {arXiv:hep-ph/0203149} \BibitemShut
  {NoStop}%
\bibitem [{\citenamefont {Tornqvist}(1995)}]{Tornqvist:1995kr}%
  \BibitemOpen
  \bibfield  {author} {\bibinfo {author} {\bibfnamefont {N.~A.}\ \bibnamefont
  {Tornqvist}},\ }\Doi {10.1007/BF01565264} {\bibfield  {journal} {\bibinfo
  {journal} {Z. Phys.},\ }\textbf {\bibinfo {volume} {C68}},\ \bibinfo {pages}
  {647} (\bibinfo {year} {1995})},\ \Eprint
  {http://arxiv.org/abs/hep-ph/9504372} {arXiv:hep-ph/9504372 [hep-ph]}
  \BibitemShut {NoStop}%
\bibitem [{\citenamefont {Kalashnikova}(2005)}]{Kalashnikova:2005ui}%
  \BibitemOpen
  \bibfield  {author} {\bibinfo {author} {\bibfnamefont {{\relax Yu}.~S.}\
  \bibnamefont {Kalashnikova}},\ }\Doi {10.1103/PhysRevD.72.034010} {\bibfield
  {journal} {\bibinfo  {journal} {Phys. Rev.},\ }\textbf {\bibinfo {volume}
  {D72}},\ \bibinfo {pages} {034010} (\bibinfo {year} {2005})},\ \Eprint
  {http://arxiv.org/abs/hep-ph/0506270} {arXiv:hep-ph/0506270 [hep-ph]}
  \BibitemShut {NoStop}%
\bibitem [{\citenamefont {Ortega}\ \emph {et~al.}(2010)\citenamefont {Ortega},
  \citenamefont {Segovia}, \citenamefont {Entem},\ and\ \citenamefont
  {Fernandez}}]{Ortega:2009hj}%
  \BibitemOpen
  \bibfield  {author} {\bibinfo {author} {\bibfnamefont {P.~G.}\ \bibnamefont
  {Ortega}}, \bibinfo {author} {\bibfnamefont {J.}~\bibnamefont {Segovia}},
  \bibinfo {author} {\bibfnamefont {D.~R.}\ \bibnamefont {Entem}}, \ and\
  \bibinfo {author} {\bibfnamefont {F.}~\bibnamefont {Fernandez}},\ }\Doi
  {10.1103/PhysRevD.81.054023} {\bibfield  {journal} {\bibinfo  {journal}
  {Phys. Rev.},\ }\textbf {\bibinfo {volume} {D81}},\ \bibinfo {pages} {054023}
  (\bibinfo {year} {2010})},\ \Eprint {http://arxiv.org/abs/0907.3997}
  {arXiv:0907.3997 [hep-ph]} \BibitemShut {NoStop}%
\bibitem [{\citenamefont {Takizawa}\ and\ \citenamefont
  {Takeuchi}(2013)}]{Takizawa:2012hy}%
  \BibitemOpen
  \bibfield  {author} {\bibinfo {author} {\bibfnamefont {M.}~\bibnamefont
  {Takizawa}}\ and\ \bibinfo {author} {\bibfnamefont {S.}~\bibnamefont
  {Takeuchi}},\ }\Doi {10.1093/ptep/ptt063} {\bibfield  {journal} {\bibinfo
  {journal} {PTEP},\ }\textbf {\bibinfo {volume} {2013}},\ \bibinfo {pages}
  {093D01} (\bibinfo {year} {2013})},\ \Eprint {http://arxiv.org/abs/1206.4877}
  {arXiv:1206.4877 [hep-ph]} \BibitemShut {NoStop}%
\bibitem [{\citenamefont {Coito}\ \emph {et~al.}(2013)\citenamefont {Coito},
  \citenamefont {Rupp},\ and\ \citenamefont {van Beveren}}]{Coito:2012vf}%
  \BibitemOpen
  \bibfield  {author} {\bibinfo {author} {\bibfnamefont {S.}~\bibnamefont
  {Coito}}, \bibinfo {author} {\bibfnamefont {G.}~\bibnamefont {Rupp}}, \ and\
  \bibinfo {author} {\bibfnamefont {E.}~\bibnamefont {van Beveren}},\ }\Doi
  {10.1140/epjc/s10052-013-2351-8} {\bibfield  {journal} {\bibinfo  {journal}
  {Eur. Phys. J.},\ }\textbf {\bibinfo {volume} {C73}},\ \bibinfo {pages}
  {2351} (\bibinfo {year} {2013})},\ \Eprint {http://arxiv.org/abs/1212.0648}
  {arXiv:1212.0648 [hep-ph]} \BibitemShut {NoStop}%
\bibitem [{\citenamefont {Sekihara}\ \emph {et~al.}(2015)\citenamefont
  {Sekihara}, \citenamefont {Hyodo},\ and\ \citenamefont
  {Jido}}]{Sekihara:2014kya}%
  \BibitemOpen
  \bibfield  {author} {\bibinfo {author} {\bibfnamefont {T.}~\bibnamefont
  {Sekihara}}, \bibinfo {author} {\bibfnamefont {T.}~\bibnamefont {Hyodo}}, \
  and\ \bibinfo {author} {\bibfnamefont {D.}~\bibnamefont {Jido}},\ }\Doi
  {10.1093/ptep/ptv081} {\bibfield  {journal} {\bibinfo  {journal} {PTEP},\
  }\textbf {\bibinfo {volume} {2015}},\ \bibinfo {pages} {063D04} (\bibinfo
  {year} {2015})},\ \Eprint {http://arxiv.org/abs/1411.2308} {arXiv:1411.2308
  [hep-ph]} \BibitemShut {NoStop}%
\bibitem [{\citenamefont {Zhou}\ and\ \citenamefont
  {Xiao}(2017)}]{Zhou:2017dwj}%
  \BibitemOpen
  \bibfield  {author} {\bibinfo {author} {\bibfnamefont {Z.-Y.}\ \bibnamefont
  {Zhou}}\ and\ \bibinfo {author} {\bibfnamefont {Z.}~\bibnamefont {Xiao}},\
  }\Doi {10.1103/PhysRevD.96.054031} {\bibfield  {journal} {\bibinfo  {journal}
  {Phys. Rev.},\ }\textbf {\bibinfo {volume} {D96}},\ \bibinfo {pages} {054031}
  (\bibinfo {year} {2017})},\ \bibinfo {note} {[Erratum: Phys. Rev. D 96,
  099905 (2017)]},\ \Eprint {http://arxiv.org/abs/1704.04438} {arXiv:1704.04438
  [hep-ph]} \BibitemShut {NoStop}%
\bibitem [{\citenamefont {Zhou}\ and\ \citenamefont
  {Xiao}(2018)}]{Zhou:2017txt}%
  \BibitemOpen
  \bibfield  {author} {\bibinfo {author} {\bibfnamefont {Z.-Y.}\ \bibnamefont
  {Zhou}}\ and\ \bibinfo {author} {\bibfnamefont {Z.}~\bibnamefont {Xiao}},\
  }\Doi {10.1103/PhysRevD.97.034011} {\bibfield  {journal} {\bibinfo  {journal}
  {Phys. Rev.},\ }\textbf {\bibinfo {volume} {D97}},\ \bibinfo {pages} {034011}
  (\bibinfo {year} {2018})},\ \Eprint {http://arxiv.org/abs/1711.01930}
  {arXiv:1711.01930 [hep-ph]} \BibitemShut {NoStop}%
\bibitem [{\citenamefont {Giacosa}\ \emph {et~al.}(2019)\citenamefont
  {Giacosa}, \citenamefont {Piotrowska},\ and\ \citenamefont
  {Coito}}]{Giacosa:2019zxw}%
  \BibitemOpen
  \bibfield  {author} {\bibinfo {author} {\bibfnamefont {F.}~\bibnamefont
  {Giacosa}}, \bibinfo {author} {\bibfnamefont {M.}~\bibnamefont {Piotrowska}},
  \ and\ \bibinfo {author} {\bibfnamefont {S.}~\bibnamefont {Coito}},\
  }\href@noop {} { (\bibinfo {year} {2019})},\ \Eprint
  {http://arxiv.org/abs/1903.06926} {arXiv:1903.06926 [hep-ph]} \BibitemShut
  {NoStop}%
\bibitem [{\citenamefont {Coito}\ and\ \citenamefont
  {Giacosa}(2019)}]{Coito:2017ppc}%
  \BibitemOpen
  \bibfield  {author} {\bibinfo {author} {\bibfnamefont {S.}~\bibnamefont
  {Coito}}\ and\ \bibinfo {author} {\bibfnamefont {F.}~\bibnamefont
  {Giacosa}},\ }\Doi {10.1016/j.nuclphysa.2018.10.083} {\bibfield  {journal}
  {\bibinfo  {journal} {Nucl. Phys. A},\ }\textbf {\bibinfo {volume} {981}},\
  \bibinfo {pages} {38} (\bibinfo {year} {2019})},\ \Eprint
  {http://arxiv.org/abs/1712.00969} {arXiv:1712.00969 [hep-ph]} \BibitemShut
  {NoStop}%
\bibitem [{\citenamefont {Godfrey}\ and\ \citenamefont
  {Isgur}(1985)}]{Godfrey:1985xj}%
  \BibitemOpen
  \bibfield  {author} {\bibinfo {author} {\bibfnamefont {S.}~\bibnamefont
  {Godfrey}}\ and\ \bibinfo {author} {\bibfnamefont {N.}~\bibnamefont
  {Isgur}},\ }\Doi {10.1103/PhysRevD.32.189} {\bibfield  {journal} {\bibinfo
  {journal} {Phys. Rev.},\ }\textbf {\bibinfo {volume} {D 32}},\ \bibinfo
  {pages} {189} (\bibinfo {year} {1985})}\BibitemShut {NoStop}%
\bibitem [{\citenamefont {Tanabashi}\ \emph {et~al.}(2018)\citenamefont
  {Tanabashi} \emph {et~al.}}]{Tanabashi:2018oca}%
  \BibitemOpen
  \bibfield  {author} {\bibinfo {author} {\bibfnamefont {M.}~\bibnamefont
  {Tanabashi}} \emph {et~al.} (\bibinfo {collaboration} {Particle Data
  Group}),\ }\Doi {10.1103/PhysRevD.98.030001} {\bibfield  {journal} {\bibinfo
  {journal} {Phys. Rev. D},\ }\textbf {\bibinfo {volume} {98}},\ \bibinfo
  {pages} {030001} (\bibinfo {year} {2018})}\BibitemShut {NoStop}%
\bibitem [{\citenamefont {Caprini}\ \emph {et~al.}(2006)\citenamefont
  {Caprini}, \citenamefont {Colangelo},\ and\ \citenamefont
  {Leutwyler}}]{Caprini:2005zr}%
  \BibitemOpen
  \bibfield  {author} {\bibinfo {author} {\bibfnamefont {I.}~\bibnamefont
  {Caprini}}, \bibinfo {author} {\bibfnamefont {G.}~\bibnamefont {Colangelo}},
  \ and\ \bibinfo {author} {\bibfnamefont {H.}~\bibnamefont {Leutwyler}},\
  }\Doi {10.1103/PhysRevLett.96.132001} {\bibfield  {journal} {\bibinfo
  {journal} {Phys.Rev.Lett.},\ }\textbf {\bibinfo {volume} {96}},\ \bibinfo
  {pages} {132001} (\bibinfo {year} {2006})},\ \Eprint
  {http://arxiv.org/abs/hep-ph/0512364} {arXiv:hep-ph/0512364 [hep-ph]}
  \BibitemShut {NoStop}%
\bibitem [{\citenamefont {Zhou}\ \emph {et~al.}(2005)\citenamefont {Zhou},
  \citenamefont {Qin}, \citenamefont {Zhang}, \citenamefont {Xiao},
  \citenamefont {Zheng} \emph {et~al.}}]{Zhou:2004ms}%
  \BibitemOpen
  \bibfield  {author} {\bibinfo {author} {\bibfnamefont {Z.}~\bibnamefont
  {Zhou}}, \bibinfo {author} {\bibfnamefont {G.}~\bibnamefont {Qin}}, \bibinfo
  {author} {\bibfnamefont {P.}~\bibnamefont {Zhang}}, \bibinfo {author}
  {\bibfnamefont {Z.}~\bibnamefont {Xiao}}, \bibinfo {author} {\bibfnamefont
  {H.}~\bibnamefont {Zheng}},  \emph {et~al.},\ }\Doi
  {10.1088/1126-6708/2005/02/043} {\bibfield  {journal} {\bibinfo  {journal}
  {JHEP},\ }\textbf {\bibinfo {volume} {0502}},\ \bibinfo {pages} {043}
  (\bibinfo {year} {2005})},\ \Eprint {http://arxiv.org/abs/hep-ph/0406271}
  {arXiv:hep-ph/0406271 [hep-ph]} \BibitemShut {NoStop}%
\bibitem [{\citenamefont {Zheng}\ \emph {et~al.}(2004)\citenamefont {Zheng},
  \citenamefont {Zhou}, \citenamefont {Qin}, \citenamefont {Xiao},
  \citenamefont {Wang},\ and\ \citenamefont {Wu}}]{Zheng:2003rw}%
  \BibitemOpen
  \bibfield  {author} {\bibinfo {author} {\bibfnamefont {H.~Q.}\ \bibnamefont
  {Zheng}}, \bibinfo {author} {\bibfnamefont {Z.~Y.}\ \bibnamefont {Zhou}},
  \bibinfo {author} {\bibfnamefont {G.~Y.}\ \bibnamefont {Qin}}, \bibinfo
  {author} {\bibfnamefont {Z.}~\bibnamefont {Xiao}}, \bibinfo {author}
  {\bibfnamefont {J.~J.}\ \bibnamefont {Wang}}, \ and\ \bibinfo {author}
  {\bibfnamefont {N.}~\bibnamefont {Wu}},\ }\Doi
  {10.1016/j.nuclphysa.2003.12.021} {\bibfield  {journal} {\bibinfo  {journal}
  {Nucl. Phys.},\ }\textbf {\bibinfo {volume} {A733}},\ \bibinfo {pages} {235}
  (\bibinfo {year} {2004})},\ \Eprint {http://arxiv.org/abs/hep-ph/0310293}
  {arXiv:hep-ph/0310293 [hep-ph]} \BibitemShut {NoStop}%
\bibitem [{\citenamefont {Descotes-Genon}\ and\ \citenamefont
  {Moussallam}(2006)}]{DescotesGenon:2006uk}%
  \BibitemOpen
  \bibfield  {author} {\bibinfo {author} {\bibfnamefont {S.}~\bibnamefont
  {Descotes-Genon}}\ and\ \bibinfo {author} {\bibfnamefont {B.}~\bibnamefont
  {Moussallam}},\ }\Doi {10.1140/epjc/s10052-006-0036-2} {\bibfield  {journal}
  {\bibinfo  {journal} {Eur. Phys. J. C},\ }\textbf {\bibinfo {volume} {48}},\
  \bibinfo {pages} {553} (\bibinfo {year} {2006})},\ \Eprint
  {http://arxiv.org/abs/hep-ph/0607133} {arXiv:hep-ph/0607133} \BibitemShut
  {NoStop}%
\bibitem [{\citenamefont {Peláez}\ and\ \citenamefont
  {Rodas}(2020)}]{Pelaez:2020uiw}%
  \BibitemOpen
  \bibfield  {author} {\bibinfo {author} {\bibfnamefont {J.}~\bibnamefont
  {Peláez}}\ and\ \bibinfo {author} {\bibfnamefont {A.}~\bibnamefont
  {Rodas}},\ }\Doi {10.1103/PhysRevLett.124.172001} {\bibfield  {journal}
  {\bibinfo  {journal} {Phys. Rev. Lett.},\ }\textbf {\bibinfo {volume}
  {124}},\ \bibinfo {pages} {172001} (\bibinfo {year} {2020})},\ \Eprint
  {http://arxiv.org/abs/2001.08153} {arXiv:2001.08153 [hep-ph]} \BibitemShut
  {NoStop}%
\bibitem [{\citenamefont {Pelaez}(2016)}]{Pelaez:2015qba}%
  \BibitemOpen
  \bibfield  {author} {\bibinfo {author} {\bibfnamefont {J.~R.}\ \bibnamefont
  {Pelaez}},\ }\Doi {10.1016/j.physrep.2016.09.001} {\bibfield  {journal}
  {\bibinfo  {journal} {Phys. Rept.},\ }\textbf {\bibinfo {volume} {658}},\
  \bibinfo {pages} {1} (\bibinfo {year} {2016})},\ \Eprint
  {http://arxiv.org/abs/1510.00653} {arXiv:1510.00653 [hep-ph]} \BibitemShut
  {NoStop}%
\bibitem [{\citenamefont {Yao}\ \emph {et~al.}(2020)\citenamefont {Yao},
  \citenamefont {Dai}, \citenamefont {Zheng},\ and\ \citenamefont
  {Zhou}}]{Yao:2020bxx}%
  \BibitemOpen
  \bibfield  {author} {\bibinfo {author} {\bibfnamefont {D.-L.}\ \bibnamefont
  {Yao}}, \bibinfo {author} {\bibfnamefont {L.-Y.}\ \bibnamefont {Dai}},
  \bibinfo {author} {\bibfnamefont {H.-Q.}\ \bibnamefont {Zheng}}, \ and\
  \bibinfo {author} {\bibfnamefont {Z.-Y.}\ \bibnamefont {Zhou}},\ }\href@noop
  {} { (\bibinfo {year} {2020})},\ \Eprint {http://arxiv.org/abs/2009.13495}
  {arXiv:2009.13495 [hep-ph]} \BibitemShut {NoStop}%
\bibitem [{\citenamefont {Close}\ and\ \citenamefont
  {Tornqvist}(2002)}]{Close:2002zu}%
  \BibitemOpen
  \bibfield  {author} {\bibinfo {author} {\bibfnamefont {F.~E.}\ \bibnamefont
  {Close}}\ and\ \bibinfo {author} {\bibfnamefont {N.~A.}\ \bibnamefont
  {Tornqvist}},\ }\Doi {10.1088/0954-3899/28/10/201} {\bibfield  {journal}
  {\bibinfo  {journal} {J. Phys. G},\ }\textbf {\bibinfo {volume} {28}},\
  \bibinfo {pages} {R249} (\bibinfo {year} {2002})},\ \Eprint
  {http://arxiv.org/abs/hep-ph/0204205} {arXiv:hep-ph/0204205} \BibitemShut
  {NoStop}%
\bibitem [{\citenamefont {Jaffe}(1977)}]{Jaffe:1976ig}%
  \BibitemOpen
  \bibfield  {author} {\bibinfo {author} {\bibfnamefont {R.~L.}\ \bibnamefont
  {Jaffe}},\ }\Doi {10.1103/PhysRevD.15.267} {\bibfield  {journal} {\bibinfo
  {journal} {Phys. Rev. D},\ }\textbf {\bibinfo {volume} {15}},\ \bibinfo
  {pages} {267} (\bibinfo {year} {1977})}\BibitemShut {NoStop}%
\bibitem [{\citenamefont {Maiani}\ \emph {et~al.}(2005)\citenamefont {Maiani},
  \citenamefont {Piccinini}, \citenamefont {Polosa},\ and\ \citenamefont
  {Riquer}}]{Maiani:2004vq}%
  \BibitemOpen
  \bibfield  {author} {\bibinfo {author} {\bibfnamefont {L.}~\bibnamefont
  {Maiani}}, \bibinfo {author} {\bibfnamefont {F.}~\bibnamefont {Piccinini}},
  \bibinfo {author} {\bibfnamefont {A.~D.}\ \bibnamefont {Polosa}}, \ and\
  \bibinfo {author} {\bibfnamefont {V.}~\bibnamefont {Riquer}},\ }\Doi
  {10.1103/PhysRevD.71.014028} {\bibfield  {journal} {\bibinfo  {journal}
  {Phys. Rev.},\ }\textbf {\bibinfo {volume} {D71}},\ \bibinfo {pages} {014028}
  (\bibinfo {year} {2005})},\ \Eprint {http://arxiv.org/abs/hep-ph/0412098}
  {arXiv:hep-ph/0412098 [hep-ph]} \BibitemShut {NoStop}%
\bibitem [{\citenamefont {Oller}\ \emph {et~al.}(1999)\citenamefont {Oller},
  \citenamefont {Oset},\ and\ \citenamefont {Pelaez}}]{Oller:1998hw}%
  \BibitemOpen
  \bibfield  {author} {\bibinfo {author} {\bibfnamefont {J.}~\bibnamefont
  {Oller}}, \bibinfo {author} {\bibfnamefont {E.}~\bibnamefont {Oset}}, \ and\
  \bibinfo {author} {\bibfnamefont {J.}~\bibnamefont {Pelaez}},\ }\Doi
  {10.1103/PhysRevD.59.074001} {\bibfield  {journal} {\bibinfo  {journal}
  {Phys. Rev. D},\ }\textbf {\bibinfo {volume} {59}},\ \bibinfo {pages}
  {074001} (\bibinfo {year} {1999})},\ \bibinfo {note} {[Erratum: Phys.Rev.D
  60, 099906 (1999), Erratum: Phys.Rev.D 75, 099903 (2007)]},\ \Eprint
  {http://arxiv.org/abs/hep-ph/9804209} {arXiv:hep-ph/9804209} \BibitemShut
  {NoStop}%
\bibitem [{\citenamefont {Guo}\ and\ \citenamefont {Oller}(2011)}]{Guo:2011pa}%
  \BibitemOpen
  \bibfield  {author} {\bibinfo {author} {\bibfnamefont {Z.-H.}\ \bibnamefont
  {Guo}}\ and\ \bibinfo {author} {\bibfnamefont {J.}~\bibnamefont {Oller}},\
  }\Doi {10.1103/PhysRevD.84.034005} {\bibfield  {journal} {\bibinfo  {journal}
  {Phys. Rev. D},\ }\textbf {\bibinfo {volume} {84}},\ \bibinfo {pages}
  {034005} (\bibinfo {year} {2011})},\ \Eprint {http://arxiv.org/abs/1104.2849}
  {arXiv:1104.2849 [hep-ph]} \BibitemShut {NoStop}%
\bibitem [{\citenamefont {Pelaez}(2004)}]{Pelaez:2003dy}%
  \BibitemOpen
  \bibfield  {author} {\bibinfo {author} {\bibfnamefont {J.}~\bibnamefont
  {Pelaez}},\ }\Doi {10.1103/PhysRevLett.92.102001} {\bibfield  {journal}
  {\bibinfo  {journal} {Phys. Rev. Lett.},\ }\textbf {\bibinfo {volume} {92}},\
  \bibinfo {pages} {102001} (\bibinfo {year} {2004})},\ \Eprint
  {http://arxiv.org/abs/hep-ph/0309292} {arXiv:hep-ph/0309292} \BibitemShut
  {NoStop}%
\bibitem [{\citenamefont {Guo}\ \emph {et~al.}(2006)\citenamefont {Guo},
  \citenamefont {Shen}, \citenamefont {Chiang}, \citenamefont {Ping},\ and\
  \citenamefont {Zou}}]{Guo:2006fu}%
  \BibitemOpen
  \bibfield  {author} {\bibinfo {author} {\bibfnamefont {F.-K.}\ \bibnamefont
  {Guo}}, \bibinfo {author} {\bibfnamefont {P.-N.}\ \bibnamefont {Shen}},
  \bibinfo {author} {\bibfnamefont {H.-C.}\ \bibnamefont {Chiang}}, \bibinfo
  {author} {\bibfnamefont {R.-G.}\ \bibnamefont {Ping}}, \ and\ \bibinfo
  {author} {\bibfnamefont {B.-S.}\ \bibnamefont {Zou}},\ }\Doi
  {10.1016/j.physletb.2006.08.064} {\bibfield  {journal} {\bibinfo  {journal}
  {Phys. Lett. B},\ }\textbf {\bibinfo {volume} {641}},\ \bibinfo {pages} {278}
  (\bibinfo {year} {2006})},\ \Eprint {http://arxiv.org/abs/hep-ph/0603072}
  {arXiv:hep-ph/0603072} \BibitemShut {NoStop}%
\bibitem [{\citenamefont {Guo}\ \emph {et~al.}(2007)\citenamefont {Guo},
  \citenamefont {Shen},\ and\ \citenamefont {Chiang}}]{Guo:2006rp}%
  \BibitemOpen
  \bibfield  {author} {\bibinfo {author} {\bibfnamefont {F.-K.}\ \bibnamefont
  {Guo}}, \bibinfo {author} {\bibfnamefont {P.-N.}\ \bibnamefont {Shen}}, \
  and\ \bibinfo {author} {\bibfnamefont {H.-C.}\ \bibnamefont {Chiang}},\ }\Doi
  {10.1016/j.physletb.2007.01.050} {\bibfield  {journal} {\bibinfo  {journal}
  {Phys. Lett. B},\ }\textbf {\bibinfo {volume} {647}},\ \bibinfo {pages} {133}
  (\bibinfo {year} {2007})},\ \Eprint {http://arxiv.org/abs/hep-ph/0610008}
  {arXiv:hep-ph/0610008} \BibitemShut {NoStop}%
\bibitem [{\citenamefont {Guo}\ \emph {et~al.}(2015)\citenamefont {Guo},
  \citenamefont {Mei{\ss}ner},\ and\ \citenamefont {Yao}}]{Guo:2015dha}%
  \BibitemOpen
  \bibfield  {author} {\bibinfo {author} {\bibfnamefont {Z.-H.}\ \bibnamefont
  {Guo}}, \bibinfo {author} {\bibfnamefont {U.-G.}\ \bibnamefont
  {Mei{\ss}ner}}, \ and\ \bibinfo {author} {\bibfnamefont {D.-L.}\ \bibnamefont
  {Yao}},\ }\Doi {10.1103/PhysRevD.92.094008} {\bibfield  {journal} {\bibinfo
  {journal} {Phys. Rev. D},\ }\textbf {\bibinfo {volume} {92}},\ \bibinfo
  {pages} {094008} (\bibinfo {year} {2015})},\ \Eprint
  {http://arxiv.org/abs/1507.03123} {arXiv:1507.03123 [hep-ph]} \BibitemShut
  {NoStop}%
\bibitem [{\citenamefont {Albaladejo}\ \emph {et~al.}(2017)\citenamefont
  {Albaladejo}, \citenamefont {Fernandez-Soler}, \citenamefont {Guo},\ and\
  \citenamefont {Nieves}}]{Albaladejo:2016lbb}%
  \BibitemOpen
  \bibfield  {author} {\bibinfo {author} {\bibfnamefont {M.}~\bibnamefont
  {Albaladejo}}, \bibinfo {author} {\bibfnamefont {P.}~\bibnamefont
  {Fernandez-Soler}}, \bibinfo {author} {\bibfnamefont {F.-K.}\ \bibnamefont
  {Guo}}, \ and\ \bibinfo {author} {\bibfnamefont {J.}~\bibnamefont {Nieves}},\
  }\Doi {10.1016/j.physletb.2017.02.036} {\bibfield  {journal} {\bibinfo
  {journal} {Phys. Lett. B},\ }\textbf {\bibinfo {volume} {767}},\ \bibinfo
  {pages} {465} (\bibinfo {year} {2017})},\ \Eprint
  {http://arxiv.org/abs/1610.06727} {arXiv:1610.06727 [hep-ph]} \BibitemShut
  {NoStop}%
\bibitem [{\citenamefont {Wu}\ \emph {et~al.}(2019)\citenamefont {Wu},
  \citenamefont {Liu}, \citenamefont {Geng}, \citenamefont {Hiyama},\ and\
  \citenamefont {Valderrama}}]{Wu:2019vsy}%
  \BibitemOpen
  \bibfield  {author} {\bibinfo {author} {\bibfnamefont {T.-W.}\ \bibnamefont
  {Wu}}, \bibinfo {author} {\bibfnamefont {M.-Z.}\ \bibnamefont {Liu}},
  \bibinfo {author} {\bibfnamefont {L.-S.}\ \bibnamefont {Geng}}, \bibinfo
  {author} {\bibfnamefont {E.}~\bibnamefont {Hiyama}}, \ and\ \bibinfo {author}
  {\bibfnamefont {M.~P.}\ \bibnamefont {Valderrama}},\ }\Doi
  {10.1103/PhysRevD.100.034029} {\bibfield  {journal} {\bibinfo  {journal}
  {Phys. Rev. D},\ }\textbf {\bibinfo {volume} {100}},\ \bibinfo {pages}
  {034029} (\bibinfo {year} {2019})},\ \Eprint
  {http://arxiv.org/abs/1906.11995} {arXiv:1906.11995 [hep-ph]} \BibitemShut
  {NoStop}%
\bibitem [{\citenamefont {van Beveren}\ \emph
  {et~al.}(1983){\natexlab{a}}\citenamefont {van Beveren}, \citenamefont
  {Rupp}, \citenamefont {Rijken},\ and\ \citenamefont
  {Dullemond}}]{vanBeveren:1982qb}%
  \BibitemOpen
  \bibfield  {author} {\bibinfo {author} {\bibfnamefont {E.}~\bibnamefont {van
  Beveren}}, \bibinfo {author} {\bibfnamefont {G.}~\bibnamefont {Rupp}},
  \bibinfo {author} {\bibfnamefont {T.}~\bibnamefont {Rijken}}, \ and\ \bibinfo
  {author} {\bibfnamefont {C.}~\bibnamefont {Dullemond}},\ }\Doi
  {10.1103/PhysRevD.27.1527} {\bibfield  {journal} {\bibinfo  {journal} {Phys.
  Rev. D},\ }\textbf {\bibinfo {volume} {27}},\ \bibinfo {pages} {1527}
  (\bibinfo {year} {1983}{\natexlab{a}})}\BibitemShut {NoStop}%
\bibitem [{\citenamefont {van Beveren}\ \emph
  {et~al.}(1983){\natexlab{b}}\citenamefont {van Beveren}, \citenamefont
  {Dullemond},\ and\ \citenamefont {Rijken}}]{vanBeveren:1983td}%
  \BibitemOpen
  \bibfield  {author} {\bibinfo {author} {\bibfnamefont {E.}~\bibnamefont {van
  Beveren}}, \bibinfo {author} {\bibfnamefont {C.}~\bibnamefont {Dullemond}}, \
  and\ \bibinfo {author} {\bibfnamefont {T.}~\bibnamefont {Rijken}},\ }\Doi
  {10.1007/BF01572256} {\bibfield  {journal} {\bibinfo  {journal} {Z.Phys.},\
  }\textbf {\bibinfo {volume} {C19}},\ \bibinfo {pages} {275} (\bibinfo {year}
  {1983}{\natexlab{b}})}\BibitemShut {NoStop}%
\bibitem [{\citenamefont {van Beveren}\ \emph {et~al.}(1986)\citenamefont {van
  Beveren}, \citenamefont {Rijken}, \citenamefont {Metzger}, \citenamefont
  {Dullemond}, \citenamefont {Rupp},\ and\ \citenamefont
  {Ribeiro}}]{vanBeveren:1986ea}%
  \BibitemOpen
  \bibfield  {author} {\bibinfo {author} {\bibfnamefont {E.}~\bibnamefont {van
  Beveren}}, \bibinfo {author} {\bibfnamefont {T.}~\bibnamefont {Rijken}},
  \bibinfo {author} {\bibfnamefont {K.}~\bibnamefont {Metzger}}, \bibinfo
  {author} {\bibfnamefont {C.}~\bibnamefont {Dullemond}}, \bibinfo {author}
  {\bibfnamefont {G.}~\bibnamefont {Rupp}}, \ and\ \bibinfo {author}
  {\bibfnamefont {J.}~\bibnamefont {Ribeiro}},\ }\Doi {10.1007/BF01571811}
  {\bibfield  {journal} {\bibinfo  {journal} {Z. Phys. C},\ }\textbf {\bibinfo
  {volume} {30}},\ \bibinfo {pages} {615} (\bibinfo {year} {1986})},\ \Eprint
  {http://arxiv.org/abs/0710.4067} {arXiv:0710.4067 [hep-ph]} \BibitemShut
  {NoStop}%
\bibitem [{\citenamefont {van Beveren}\ \emph {et~al.}(2006)\citenamefont {van
  Beveren}, \citenamefont {Bugg}, \citenamefont {Kleefeld},\ and\ \citenamefont
  {Rupp}}]{vanBeveren:2006ua}%
  \BibitemOpen
  \bibfield  {author} {\bibinfo {author} {\bibfnamefont {E.}~\bibnamefont {van
  Beveren}}, \bibinfo {author} {\bibfnamefont {D.}~\bibnamefont {Bugg}},
  \bibinfo {author} {\bibfnamefont {F.}~\bibnamefont {Kleefeld}}, \ and\
  \bibinfo {author} {\bibfnamefont {G.}~\bibnamefont {Rupp}},\ }\Doi
  {10.1016/j.physletb.2006.08.051} {\bibfield  {journal} {\bibinfo  {journal}
  {Phys. Lett. B},\ }\textbf {\bibinfo {volume} {641}},\ \bibinfo {pages} {265}
  (\bibinfo {year} {2006})},\ \Eprint {http://arxiv.org/abs/hep-ph/0606022}
  {arXiv:hep-ph/0606022} \BibitemShut {NoStop}%
\bibitem [{\citenamefont {van Beveren}\ and\ \citenamefont
  {Rupp}(2003)}]{vanBeveren:2003kd}%
  \BibitemOpen
  \bibfield  {author} {\bibinfo {author} {\bibfnamefont {E.}~\bibnamefont {van
  Beveren}}\ and\ \bibinfo {author} {\bibfnamefont {G.}~\bibnamefont {Rupp}},\
  }\Doi {10.1103/PhysRevLett.91.012003} {\bibfield  {journal} {\bibinfo
  {journal} {Phys. Rev. Lett.},\ }\textbf {\bibinfo {volume} {91}},\ \bibinfo
  {pages} {012003} (\bibinfo {year} {2003})},\ \Eprint
  {http://arxiv.org/abs/hep-ph/0305035} {arXiv:hep-ph/0305035} \BibitemShut
  {NoStop}%
\bibitem [{\citenamefont {Heikkila}\ \emph {et~al.}(1984)\citenamefont
  {Heikkila}, \citenamefont {Tornqvist},\ and\ \citenamefont
  {Ono}}]{Heikkila:1983wd}%
  \BibitemOpen
  \bibfield  {author} {\bibinfo {author} {\bibfnamefont {K.}~\bibnamefont
  {Heikkila}}, \bibinfo {author} {\bibfnamefont {N.~A.}\ \bibnamefont
  {Tornqvist}}, \ and\ \bibinfo {author} {\bibfnamefont {S.}~\bibnamefont
  {Ono}},\ }\Doi {10.1103/PhysRevD.29.110} {\bibfield  {journal} {\bibinfo
  {journal} {Phys. Rev.},\ }\textbf {\bibinfo {volume} {D 29}},\ \bibinfo
  {pages} {110} (\bibinfo {year} {1984})}\BibitemShut {NoStop}%
\bibitem [{\citenamefont {Zhou}\ and\ \citenamefont
  {Xiao}(2011){\natexlab{a}}}]{Zhou:2010ra}%
  \BibitemOpen
  \bibfield  {author} {\bibinfo {author} {\bibfnamefont {Z.-Y.}\ \bibnamefont
  {Zhou}}\ and\ \bibinfo {author} {\bibfnamefont {Z.}~\bibnamefont {Xiao}},\
  }\Doi {10.1103/PhysRevD.83.014010} {\bibfield  {journal} {\bibinfo  {journal}
  {Phys. Rev.},\ }\textbf {\bibinfo {volume} {D 83}},\ \bibinfo {pages}
  {014010} (\bibinfo {year} {2011}{\natexlab{a}})},\ \Eprint
  {http://arxiv.org/abs/1007.2072} {arXiv:1007.2072 [hep-ph]} \BibitemShut
  {NoStop}%
\bibitem [{\citenamefont {Boglione}\ and\ \citenamefont
  {Pennington}(1997)}]{Boglione:1997aw}%
  \BibitemOpen
  \bibfield  {author} {\bibinfo {author} {\bibfnamefont {M.}~\bibnamefont
  {Boglione}}\ and\ \bibinfo {author} {\bibfnamefont {M.~R.}\ \bibnamefont
  {Pennington}},\ }\Doi {10.1103/PhysRevLett.79.1998} {\bibfield  {journal}
  {\bibinfo  {journal} {Phys. Rev. Lett.},\ }\textbf {\bibinfo {volume} {79}},\
  \bibinfo {pages} {1998} (\bibinfo {year} {1997})},\ \Eprint
  {http://arxiv.org/abs/hep-ph/9703257} {arXiv:hep-ph/9703257} \BibitemShut
  {NoStop}%
\bibitem [{\citenamefont {Geiger}\ and\ \citenamefont
  {Isgur}(1993)}]{Geiger:1992va}%
  \BibitemOpen
  \bibfield  {author} {\bibinfo {author} {\bibfnamefont {P.}~\bibnamefont
  {Geiger}}\ and\ \bibinfo {author} {\bibfnamefont {N.}~\bibnamefont {Isgur}},\
  }\Doi {10.1103/PhysRevD.47.5050} {\bibfield  {journal} {\bibinfo  {journal}
  {Phys. Rev. D},\ }\textbf {\bibinfo {volume} {47}},\ \bibinfo {pages} {5050}
  (\bibinfo {year} {1993})}\BibitemShut {NoStop}%
\bibitem [{\citenamefont {Badalian}\ \emph {et~al.}(2020)\citenamefont
  {Badalian}, \citenamefont {Lukashov},\ and\ \citenamefont
  {Simonov}}]{Badalian:2020wua}%
  \BibitemOpen
  \bibfield  {author} {\bibinfo {author} {\bibfnamefont {A.}~\bibnamefont
  {Badalian}}, \bibinfo {author} {\bibfnamefont {M.}~\bibnamefont {Lukashov}},
  \ and\ \bibinfo {author} {\bibfnamefont {Y.~A.}\ \bibnamefont {Simonov}},\
  }\href@noop {} { (\bibinfo {year} {2020})},\ \Eprint
  {http://arxiv.org/abs/2001.07113} {arXiv:2001.07113 [hep-ph]} \BibitemShut
  {NoStop}%
\bibitem [{\citenamefont {Wolkanowski}\ \emph
  {et~al.}(2016){\natexlab{a}}\citenamefont {Wolkanowski}, \citenamefont
  {So\l{}tysiak},\ and\ \citenamefont {Giacosa}}]{Wolkanowski:2015jtc}%
  \BibitemOpen
  \bibfield  {author} {\bibinfo {author} {\bibfnamefont {T.}~\bibnamefont
  {Wolkanowski}}, \bibinfo {author} {\bibfnamefont {M.}~\bibnamefont
  {So\l{}tysiak}}, \ and\ \bibinfo {author} {\bibfnamefont {F.}~\bibnamefont
  {Giacosa}},\ }\Doi {10.1016/j.nuclphysb.2016.05.025} {\bibfield  {journal}
  {\bibinfo  {journal} {Nucl. Phys. B},\ }\textbf {\bibinfo {volume} {909}},\
  \bibinfo {pages} {418} (\bibinfo {year} {2016}{\natexlab{a}})},\ \Eprint
  {http://arxiv.org/abs/1512.01071} {arXiv:1512.01071 [hep-ph]} \BibitemShut
  {NoStop}%
\bibitem [{\citenamefont {Wolkanowski}\ \emph
  {et~al.}(2016){\natexlab{b}}\citenamefont {Wolkanowski}, \citenamefont
  {Giacosa},\ and\ \citenamefont {Rischke}}]{Wolkanowski:2015lsa}%
  \BibitemOpen
  \bibfield  {author} {\bibinfo {author} {\bibfnamefont {T.}~\bibnamefont
  {Wolkanowski}}, \bibinfo {author} {\bibfnamefont {F.}~\bibnamefont
  {Giacosa}}, \ and\ \bibinfo {author} {\bibfnamefont {D.~H.}\ \bibnamefont
  {Rischke}},\ }\Doi {10.1103/PhysRevD.93.014002} {\bibfield  {journal}
  {\bibinfo  {journal} {Phys. Rev. D},\ }\textbf {\bibinfo {volume} {93}},\
  \bibinfo {pages} {014002} (\bibinfo {year} {2016}{\natexlab{b}})},\ \Eprint
  {http://arxiv.org/abs/1508.00372} {arXiv:1508.00372 [hep-ph]} \BibitemShut
  {NoStop}%
\bibitem [{\citenamefont {Zhou}\ and\ \citenamefont
  {Xiao}(2020)}]{Zhou:2020vnz}%
  \BibitemOpen
  \bibfield  {author} {\bibinfo {author} {\bibfnamefont {Z.-Y.}\ \bibnamefont
  {Zhou}}\ and\ \bibinfo {author} {\bibfnamefont {Z.}~\bibnamefont {Xiao}},\
  }\Doi {10.1140/epjc/s10052-020-08752-8} {\bibfield  {journal} {\bibinfo
  {journal} {Eur. Phys. J. C},\ }\textbf {\bibinfo {volume} {80}},\ \bibinfo
  {pages} {1191} (\bibinfo {year} {2020})},\ \Eprint
  {http://arxiv.org/abs/2008.02684} {arXiv:2008.02684 [hep-ph]} \BibitemShut
  {NoStop}%
\bibitem [{\citenamefont {Antoniou}\ \emph {et~al.}(1998)\citenamefont
  {Antoniou}, \citenamefont {Gadella}, \citenamefont {Prigogine},\ and\
  \citenamefont {Pronko}}]{Antoniou:1998JMP}%
  \BibitemOpen
  \bibfield  {author} {\bibinfo {author} {\bibfnamefont {I.}~\bibnamefont
  {Antoniou}}, \bibinfo {author} {\bibfnamefont {M.}~\bibnamefont {Gadella}},
  \bibinfo {author} {\bibfnamefont {I.}~\bibnamefont {Prigogine}}, \ and\
  \bibinfo {author} {\bibfnamefont {G.~P.}\ \bibnamefont {Pronko}},\ }\Doi
  {http://dx.doi.org/10.1063/1.532235} {\bibfield  {journal} {\bibinfo
  {journal} {J. Math. Phys.},\ }\textbf {\bibinfo {volume} {39}},\ \bibinfo
  {pages} {2995} (\bibinfo {year} {1998})}\BibitemShut {NoStop}%
\bibitem [{\citenamefont {Fuda}(2012)}]{Fuda:2012xd}%
  \BibitemOpen
  \bibfield  {author} {\bibinfo {author} {\bibfnamefont {M.~G.}\ \bibnamefont
  {Fuda}},\ }\Doi {10.1103/PhysRevC.86.055205} {\bibfield  {journal} {\bibinfo
  {journal} {Phys. Rev. C},\ }\textbf {\bibinfo {volume} {86}},\ \bibinfo
  {pages} {055205} (\bibinfo {year} {2012})}\BibitemShut {NoStop}%
\bibitem [{\citenamefont {Friedrichs}(1948)}]{Friedrichs:1948}%
  \BibitemOpen
  \bibfield  {author} {\bibinfo {author} {\bibfnamefont {K.~O.}\ \bibnamefont
  {Friedrichs}},\ }\Doi {10.1002/cpa.3160010404} {\bibfield  {journal}
  {\bibinfo  {journal} {Commun. Pure Appl. Math.},\ }\textbf {\bibinfo {volume}
  {1}},\ \bibinfo {pages} {361} (\bibinfo {year} {1948})}\BibitemShut {NoStop}%
\bibitem [{\citenamefont {Lee}(1954)}]{Lee:1954iq}%
  \BibitemOpen
  \bibfield  {author} {\bibinfo {author} {\bibfnamefont {T.~D.}\ \bibnamefont
  {Lee}},\ }\Doi {10.1103/PhysRev.95.1329} {\bibfield  {journal} {\bibinfo
  {journal} {Phys. Rev.},\ }\textbf {\bibinfo {volume} {95}},\ \bibinfo {pages}
  {1329} (\bibinfo {year} {1954})}\BibitemShut {NoStop}%
\bibitem [{\citenamefont {Facchi}\ \emph {et~al.}(2001)\citenamefont {Facchi},
  \citenamefont {Nakazato},\ and\ \citenamefont
  {Pascazio}}]{PhysRevLett.86.2699}%
  \BibitemOpen
  \bibfield  {author} {\bibinfo {author} {\bibfnamefont {P.}~\bibnamefont
  {Facchi}}, \bibinfo {author} {\bibfnamefont {H.}~\bibnamefont {Nakazato}}, \
  and\ \bibinfo {author} {\bibfnamefont {S.}~\bibnamefont {Pascazio}},\ }\Doi
  {10.1103/PhysRevLett.86.2699} {\bibfield  {journal} {\bibinfo  {journal}
  {Phys. Rev. Lett.},\ }\textbf {\bibinfo {volume} {86}},\ \bibinfo {pages}
  {2699} (\bibinfo {year} {2001})}\BibitemShut {NoStop}%
\bibitem [{\citenamefont {Jaynes}\ and\ \citenamefont
  {Cummings}(1963)}]{JaynesCummings}%
  \BibitemOpen
  \bibfield  {author} {\bibinfo {author} {\bibfnamefont {E.~T.}\ \bibnamefont
  {Jaynes}}\ and\ \bibinfo {author} {\bibfnamefont {F.~W.}\ \bibnamefont
  {Cummings}},\ }\href@noop {} {\bibfield  {journal} {\bibinfo  {journal}
  {Proc. IEEE},\ }\textbf {\bibinfo {volume} {51}},\ \bibinfo {pages} {89}
  (\bibinfo {year} {1963})}\BibitemShut {NoStop}%
\bibitem [{\citenamefont {Giacosa}(2012)}]{Giacosa:2011xa}%
  \BibitemOpen
  \bibfield  {author} {\bibinfo {author} {\bibfnamefont {F.}~\bibnamefont
  {Giacosa}},\ }\Doi {10.1007/s10701-012-9667-3} {\bibfield  {journal}
  {\bibinfo  {journal} {Found. Phys.},\ }\textbf {\bibinfo {volume} {42}},\
  \bibinfo {pages} {1262} (\bibinfo {year} {2012})},\ \Eprint
  {http://arxiv.org/abs/1110.5923} {arXiv:1110.5923 [nucl-th]} \BibitemShut
  {NoStop}%
\bibitem [{\citenamefont {Liu}\ \emph {et~al.}(2016)\citenamefont {Liu},
  \citenamefont {Kamleh}, \citenamefont {Leinweber}, \citenamefont {Stokes},
  \citenamefont {Thomas},\ and\ \citenamefont {Wu}}]{Liu:2015ktc}%
  \BibitemOpen
  \bibfield  {author} {\bibinfo {author} {\bibfnamefont {Z.-W.}\ \bibnamefont
  {Liu}}, \bibinfo {author} {\bibfnamefont {W.}~\bibnamefont {Kamleh}},
  \bibinfo {author} {\bibfnamefont {D.~B.}\ \bibnamefont {Leinweber}}, \bibinfo
  {author} {\bibfnamefont {F.~M.}\ \bibnamefont {Stokes}}, \bibinfo {author}
  {\bibfnamefont {A.~W.}\ \bibnamefont {Thomas}}, \ and\ \bibinfo {author}
  {\bibfnamefont {J.-J.}\ \bibnamefont {Wu}},\ }\Doi
  {10.1103/PhysRevLett.116.082004} {\bibfield  {journal} {\bibinfo  {journal}
  {Phys. Rev. Lett.},\ }\textbf {\bibinfo {volume} {116}},\ \bibinfo {pages}
  {082004} (\bibinfo {year} {2016})},\ \Eprint
  {http://arxiv.org/abs/1512.00140} {arXiv:1512.00140 [hep-lat]} \BibitemShut
  {NoStop}%
\bibitem [{\citenamefont {Lo}\ and\ \citenamefont
  {Giacosa}(2019)}]{Lo:2019who}%
  \BibitemOpen
  \bibfield  {author} {\bibinfo {author} {\bibfnamefont {P.~M.}\ \bibnamefont
  {Lo}}\ and\ \bibinfo {author} {\bibfnamefont {F.}~\bibnamefont {Giacosa}},\
  }\Doi {10.1140/epjc/s10052-019-6844-y} {\bibfield  {journal} {\bibinfo
  {journal} {Eur. Phys. J. C},\ }\textbf {\bibinfo {volume} {79}},\ \bibinfo
  {pages} {336} (\bibinfo {year} {2019})},\ \Eprint
  {http://arxiv.org/abs/1902.03203} {arXiv:1902.03203 [hep-ph]} \BibitemShut
  {NoStop}%
\bibitem [{\citenamefont {Xiao}\ and\ \citenamefont
  {Zhou}(2016)}]{Xiao:2016dsx}%
  \BibitemOpen
  \bibfield  {author} {\bibinfo {author} {\bibfnamefont {Z.}~\bibnamefont
  {Xiao}}\ and\ \bibinfo {author} {\bibfnamefont {Z.-Y.}\ \bibnamefont
  {Zhou}},\ }\Doi {10.1103/PhysRevD.94.076006} {\bibfield  {journal} {\bibinfo
  {journal} {Phys. Rev.},\ }\textbf {\bibinfo {volume} {D 94}},\ \bibinfo
  {pages} {076006} (\bibinfo {year} {2016})},\ \Eprint
  {http://arxiv.org/abs/1608.00468} {arXiv:1608.00468 [hep-ph]} \BibitemShut
  {NoStop}%
\bibitem [{\citenamefont {Xiao}\ and\ \citenamefont
  {Zhou}(2017)}]{Xiao:2016mon}%
  \BibitemOpen
  \bibfield  {author} {\bibinfo {author} {\bibfnamefont {Z.}~\bibnamefont
  {Xiao}}\ and\ \bibinfo {author} {\bibfnamefont {Z.-Y.}\ \bibnamefont
  {Zhou}},\ }\Doi {10.1063/1.4993193} {\bibfield  {journal} {\bibinfo
  {journal} {J. Math. Phys.},\ }\textbf {\bibinfo {volume} {58}},\ \bibinfo
  {pages} {072102} (\bibinfo {year} {2017})},\ \Eprint
  {http://arxiv.org/abs/1610.07460} {arXiv:1610.07460 [hep-ph]} \BibitemShut
  {NoStop}%
\bibitem [{\citenamefont {Dosch}\ and\ \citenamefont
  {Gromes}(1986)}]{Dosch:1986dp}%
  \BibitemOpen
  \bibfield  {author} {\bibinfo {author} {\bibfnamefont {H.~G.}\ \bibnamefont
  {Dosch}}\ and\ \bibinfo {author} {\bibfnamefont {D.}~\bibnamefont {Gromes}},\
  }\Doi {10.1103/PhysRevD.33.1378} {\bibfield  {journal} {\bibinfo  {journal}
  {Phys. Rev. D},\ }\textbf {\bibinfo {volume} {33}},\ \bibinfo {pages} {1378}
  (\bibinfo {year} {1986})}\BibitemShut {NoStop}%
\bibitem [{\citenamefont {Kokoski}\ and\ \citenamefont
  {Isgur}(1987)}]{Kokoski:1985is}%
  \BibitemOpen
  \bibfield  {author} {\bibinfo {author} {\bibfnamefont {R.}~\bibnamefont
  {Kokoski}}\ and\ \bibinfo {author} {\bibfnamefont {N.}~\bibnamefont
  {Isgur}},\ }\Doi {10.1103/PhysRevD.35.907} {\bibfield  {journal} {\bibinfo
  {journal} {Phys. Rev.},\ }\textbf {\bibinfo {volume} {D 35}},\ \bibinfo
  {pages} {907} (\bibinfo {year} {1987})}\BibitemShut {NoStop}%
\bibitem [{\citenamefont {Ackleh}\ \emph {et~al.}(1996)\citenamefont {Ackleh},
  \citenamefont {Barnes},\ and\ \citenamefont {Swanson}}]{Ackleh:1996yt}%
  \BibitemOpen
  \bibfield  {author} {\bibinfo {author} {\bibfnamefont {E.~S.}\ \bibnamefont
  {Ackleh}}, \bibinfo {author} {\bibfnamefont {T.}~\bibnamefont {Barnes}}, \
  and\ \bibinfo {author} {\bibfnamefont {E.~S.}\ \bibnamefont {Swanson}},\
  }\Doi {10.1103/PhysRevD.54.6811} {\bibfield  {journal} {\bibinfo  {journal}
  {Phys. Rev.},\ }\textbf {\bibinfo {volume} {D 54}},\ \bibinfo {pages} {6811}
  (\bibinfo {year} {1996})},\ \Eprint {http://arxiv.org/abs/hep-ph/9604355}
  {arXiv:hep-ph/9604355 [hep-ph]} \BibitemShut {NoStop}%
\bibitem [{\citenamefont {Micu}(1969)}]{Micu:1968mk}%
  \BibitemOpen
  \bibfield  {author} {\bibinfo {author} {\bibfnamefont {L.}~\bibnamefont
  {Micu}},\ }\Doi {10.1016/0550-3213(69)90039-X} {\bibfield  {journal}
  {\bibinfo  {journal} {Nucl. Phys.},\ }\textbf {\bibinfo {volume} {B10}},\
  \bibinfo {pages} {521} (\bibinfo {year} {1969})}\BibitemShut {NoStop}%
\bibitem [{\citenamefont {Le~Yaouanc}\ \emph {et~al.}(1973)\citenamefont
  {Le~Yaouanc}, \citenamefont {Oliver}, \citenamefont {Pene},\ and\
  \citenamefont {Raynal}}]{LeYaouanc:1972vsx}%
  \BibitemOpen
  \bibfield  {author} {\bibinfo {author} {\bibfnamefont {A.}~\bibnamefont
  {Le~Yaouanc}}, \bibinfo {author} {\bibfnamefont {L.}~\bibnamefont {Oliver}},
  \bibinfo {author} {\bibfnamefont {O.}~\bibnamefont {Pene}}, \ and\ \bibinfo
  {author} {\bibfnamefont {J.}~\bibnamefont {Raynal}},\ }\Doi
  {10.1103/PhysRevD.8.2223} {\bibfield  {journal} {\bibinfo  {journal} {Phys.
  Rev. D},\ }\textbf {\bibinfo {volume} {8}},\ \bibinfo {pages} {2223}
  (\bibinfo {year} {1973})}\BibitemShut {NoStop}%
\bibitem [{\citenamefont {Blundell}\ and\ \citenamefont
  {Godfrey}(1996)}]{Blundell:1995ev}%
  \BibitemOpen
  \bibfield  {author} {\bibinfo {author} {\bibfnamefont {H.~G.}\ \bibnamefont
  {Blundell}}\ and\ \bibinfo {author} {\bibfnamefont {S.}~\bibnamefont
  {Godfrey}},\ }\Doi {10.1103/PhysRevD.53.3700} {\bibfield  {journal} {\bibinfo
   {journal} {Phys. Rev.},\ }\textbf {\bibinfo {volume} {D 53}},\ \bibinfo
  {pages} {3700} (\bibinfo {year} {1996})},\ \Eprint
  {http://arxiv.org/abs/hep-ph/9508264} {arXiv:hep-ph/9508264 [hep-ph]}
  \BibitemShut {NoStop}%
\bibitem [{\citenamefont {Macfarlane}(1963)}]{macfarlane1963}%
  \BibitemOpen
  \bibfield  {author} {\bibinfo {author} {\bibfnamefont {A.~J.}\ \bibnamefont
  {Macfarlane}},\ }\Doi {10.1063/1.1703981} {\bibfield  {journal} {\bibinfo
  {journal} {Journal of Mathematical Physics},\ }\textbf {\bibinfo {volume}
  {4}},\ \bibinfo {pages} {490} (\bibinfo {year} {1963})},\ \Eprint
  {http://arxiv.org/abs/https://doi.org/10.1063/1.1703981}
  {https://doi.org/10.1063/1.1703981} \BibitemShut {NoStop}%
\bibitem [{\citenamefont {McKerrell}(1964)}]{McKerrell1964}%
  \BibitemOpen
  \bibfield  {author} {\bibinfo {author} {\bibfnamefont {A.}~\bibnamefont
  {McKerrell}},\ }\Doi {10.1007/BF02748855} {\bibfield  {journal} {\bibinfo
  {journal} {Nuovo Cimento},\ }\textbf {\bibinfo {volume} {34}},\ \bibinfo
  {pages} {1289} (\bibinfo {year} {1964})}\BibitemShut {NoStop}%
\bibitem [{\citenamefont {Jacob}\ and\ \citenamefont
  {Wick}(1959)}]{Jacob:1959at}%
  \BibitemOpen
  \bibfield  {author} {\bibinfo {author} {\bibfnamefont {M.}~\bibnamefont
  {Jacob}}\ and\ \bibinfo {author} {\bibfnamefont {G.~C.}\ \bibnamefont
  {Wick}},\ }\Doi {10.1016/0003-4916(59)90051-X} {\bibfield  {journal}
  {\bibinfo  {journal} {Annals Phys.},\ }\textbf {\bibinfo {volume} {7}},\
  \bibinfo {pages} {404} (\bibinfo {year} {1959})},\ \bibinfo {note} {[Annals
  Phys.281,774(2000)]}\BibitemShut {NoStop}%
\bibitem [{\citenamefont {Zhou}\ and\ \citenamefont
  {Xiao}(2011){\natexlab{b}}}]{Zhou:2011sp}%
  \BibitemOpen
  \bibfield  {author} {\bibinfo {author} {\bibfnamefont {Z.-Y.}\ \bibnamefont
  {Zhou}}\ and\ \bibinfo {author} {\bibfnamefont {Z.}~\bibnamefont {Xiao}},\
  }\Doi {10.1103/PhysRevD.84.034023} {\bibfield  {journal} {\bibinfo  {journal}
  {Phys. Rev.},\ }\textbf {\bibinfo {volume} {D 84}},\ \bibinfo {pages}
  {034023} (\bibinfo {year} {2011}{\natexlab{b}})},\ \Eprint
  {http://arxiv.org/abs/1105.6025} {arXiv:1105.6025 [hep-ph]} \BibitemShut
  {NoStop}%
\bibitem [{\citenamefont {Deandrea}\ \emph {et~al.}(2001)\citenamefont
  {Deandrea}, \citenamefont {Gatto}, \citenamefont {Nardulli}, \citenamefont
  {Polosa},\ and\ \citenamefont {Tornqvist}}]{Deandrea:2000yc}%
  \BibitemOpen
  \bibfield  {author} {\bibinfo {author} {\bibfnamefont {A.}~\bibnamefont
  {Deandrea}}, \bibinfo {author} {\bibfnamefont {R.}~\bibnamefont {Gatto}},
  \bibinfo {author} {\bibfnamefont {G.}~\bibnamefont {Nardulli}}, \bibinfo
  {author} {\bibfnamefont {A.}~\bibnamefont {Polosa}}, \ and\ \bibinfo {author}
  {\bibfnamefont {N.}~\bibnamefont {Tornqvist}},\ }\Doi
  {10.1016/S0370-2693(01)00183-6} {\bibfield  {journal} {\bibinfo  {journal}
  {Phys. Lett. B},\ }\textbf {\bibinfo {volume} {502}},\ \bibinfo {pages} {79}
  (\bibinfo {year} {2001})},\ \Eprint {http://arxiv.org/abs/hep-ph/0012120}
  {arXiv:hep-ph/0012120} \BibitemShut {NoStop}%
\bibitem [{\citenamefont {Acciarri}\ \emph {et~al.}(2001)\citenamefont
  {Acciarri} \emph {et~al.}}]{Acciarri:2000ex}%
  \BibitemOpen
  \bibfield  {author} {\bibinfo {author} {\bibfnamefont {M.}~\bibnamefont
  {Acciarri}} \emph {et~al.} (\bibinfo {collaboration} {L3}),\ }\Doi
  {10.1016/S0370-2693(01)00116-2} {\bibfield  {journal} {\bibinfo  {journal}
  {Phys. Lett. B},\ }\textbf {\bibinfo {volume} {501}},\ \bibinfo {pages} {173}
  (\bibinfo {year} {2001})},\ \Eprint {http://arxiv.org/abs/hep-ex/0011037}
  {arXiv:hep-ex/0011037} \BibitemShut {NoStop}%
\bibitem [{\citenamefont {Flatte}(1976)}]{Flatte:1976xu}%
  \BibitemOpen
  \bibfield  {author} {\bibinfo {author} {\bibfnamefont {S.~M.}\ \bibnamefont
  {Flatte}},\ }\Doi {10.1016/0370-2693(76)90654-7} {\bibfield  {journal}
  {\bibinfo  {journal} {Phys. Lett. B},\ }\textbf {\bibinfo {volume} {63}},\
  \bibinfo {pages} {224} (\bibinfo {year} {1976})}\BibitemShut {NoStop}%
\bibitem [{\citenamefont {Mei{\ss}ner}(2020)}]{Meissner:2020khl}%
  \BibitemOpen
  \bibfield  {author} {\bibinfo {author} {\bibfnamefont {U.-G.}\ \bibnamefont
  {Mei{\ss}ner}},\ }\Doi {10.3390/sym12060981} {\bibfield  {journal} {\bibinfo
  {journal} {Symmetry},\ }\textbf {\bibinfo {volume} {12}},\ \bibinfo {pages}
  {981} (\bibinfo {year} {2020})},\ \Eprint {http://arxiv.org/abs/2005.06909}
  {arXiv:2005.06909 [hep-ph]} \BibitemShut {NoStop}%
\bibitem [{\citenamefont {Moir}\ \emph {et~al.}(2016)\citenamefont {Moir},
  \citenamefont {Peardon}, \citenamefont {Ryan}, \citenamefont {Thomas},\ and\
  \citenamefont {Wilson}}]{Moir:2016srx}%
  \BibitemOpen
  \bibfield  {author} {\bibinfo {author} {\bibfnamefont {G.}~\bibnamefont
  {Moir}}, \bibinfo {author} {\bibfnamefont {M.}~\bibnamefont {Peardon}},
  \bibinfo {author} {\bibfnamefont {S.~M.}\ \bibnamefont {Ryan}}, \bibinfo
  {author} {\bibfnamefont {C.~E.}\ \bibnamefont {Thomas}}, \ and\ \bibinfo
  {author} {\bibfnamefont {D.~J.}\ \bibnamefont {Wilson}},\ }\Doi
  {10.1007/JHEP10(2016)011} {\bibfield  {journal} {\bibinfo  {journal} {JHEP},\
  }\textbf {\bibinfo {volume} {10}},\ \bibinfo {pages} {011} (\bibinfo {year}
  {2016})},\ \Eprint {http://arxiv.org/abs/1607.07093} {arXiv:1607.07093
  [hep-lat]} \BibitemShut {NoStop}%
\bibitem [{Note1()}]{Note1}%
  \BibitemOpen
  \bibinfo {note} {A frequently used parameterization of an $S$-wave resonance
  contribution to the $S$-matrix is $S=\protect \frac {s-M^2-i\rho (s)G_1
  }{s-M^2+i\rho (s)G_1 }$ which contributes $180^\circ $ to the phase shift,
  where $\rho (s)=2k/\protect \sqrt s$ is the kinematic factor, $G_1$ being a
  constant. However, this representation only works for narrow resonances,
  since besides a pair of resonance poles on the second Riemann sheet, there is
  also a virtual state pole. A more suitable parametrization is $S=\protect
  \frac {s-M^2-is\protect \,\rho (s)G_2 }{s-M^2+is\protect \,\rho (s)G_2 }$
  which contributes less than $180^\circ $ (see \cite {Zheng:2003cr} for
  details).}\BibitemShut {Stop}%
\bibitem [{\citenamefont {Mercer}\ \emph {et~al.}(1971)\citenamefont {Mercer}
  \emph {et~al.}}]{Mercer:1971kn}%
  \BibitemOpen
  \bibfield  {author} {\bibinfo {author} {\bibfnamefont {R.}~\bibnamefont
  {Mercer}} \emph {et~al.},\ }\Doi {10.1016/0550-3213(71)90483-4} {\bibfield
  {journal} {\bibinfo  {journal} {Nucl. Phys. B},\ }\textbf {\bibinfo {volume}
  {32}},\ \bibinfo {pages} {381} (\bibinfo {year} {1971})}\BibitemShut
  {NoStop}%
\bibitem [{\citenamefont {Estabrooks}\ \emph {et~al.}(1978)\citenamefont
  {Estabrooks}, \citenamefont {Carnegie}, \citenamefont {Martin}, \citenamefont
  {Dunwoodie}, \citenamefont {Lasinski},\ and\ \citenamefont
  {Leith}}]{Estabrooks:1977xe}%
  \BibitemOpen
  \bibfield  {author} {\bibinfo {author} {\bibfnamefont {P.}~\bibnamefont
  {Estabrooks}}, \bibinfo {author} {\bibfnamefont {R.}~\bibnamefont
  {Carnegie}}, \bibinfo {author} {\bibfnamefont {A.~D.}\ \bibnamefont
  {Martin}}, \bibinfo {author} {\bibfnamefont {W.}~\bibnamefont {Dunwoodie}},
  \bibinfo {author} {\bibfnamefont {T.}~\bibnamefont {Lasinski}}, \ and\
  \bibinfo {author} {\bibfnamefont {D.~W.}\ \bibnamefont {Leith}},\ }\Doi
  {10.1016/0550-3213(78)90238-9} {\bibfield  {journal} {\bibinfo  {journal}
  {Nucl. Phys. B},\ }\textbf {\bibinfo {volume} {133}},\ \bibinfo {pages} {490}
  (\bibinfo {year} {1978})}\BibitemShut {NoStop}%
\bibitem [{\citenamefont {Bingham}\ \emph {et~al.}(1972)\citenamefont {Bingham}
  \emph {et~al.}}]{Bingham:1972vy}%
  \BibitemOpen
  \bibfield  {author} {\bibinfo {author} {\bibfnamefont {H.}~\bibnamefont
  {Bingham}} \emph {et~al.},\ }\Doi {10.1016/0550-3213(72)90419-1} {\bibfield
  {journal} {\bibinfo  {journal} {Nucl. Phys. B},\ }\textbf {\bibinfo {volume}
  {41}},\ \bibinfo {pages} {1} (\bibinfo {year} {1972})}\BibitemShut {NoStop}%
\bibitem [{\citenamefont {Aston}\ \emph {et~al.}(1988)\citenamefont {Aston}
  \emph {et~al.}}]{Aston:1987ir}%
  \BibitemOpen
  \bibfield  {author} {\bibinfo {author} {\bibfnamefont {D.}~\bibnamefont
  {Aston}} \emph {et~al.},\ }\Doi {10.1016/0550-3213(88)90028-4} {\bibfield
  {journal} {\bibinfo  {journal} {Nucl. Phys. B},\ }\textbf {\bibinfo {volume}
  {296}},\ \bibinfo {pages} {493} (\bibinfo {year} {1988})}\BibitemShut
  {NoStop}%
\bibitem [{\citenamefont {Dobado}\ and\ \citenamefont
  {Pelaez}(1997)}]{Dobado:1996ps}%
  \BibitemOpen
  \bibfield  {author} {\bibinfo {author} {\bibfnamefont {A.}~\bibnamefont
  {Dobado}}\ and\ \bibinfo {author} {\bibfnamefont {J.~R.}\ \bibnamefont
  {Pelaez}},\ }\Doi {10.1103/PhysRevD.56.3057} {\bibfield  {journal} {\bibinfo
  {journal} {Phys. Rev. D},\ }\textbf {\bibinfo {volume} {56}},\ \bibinfo
  {pages} {3057} (\bibinfo {year} {1997})},\ \Eprint
  {http://arxiv.org/abs/hep-ph/9604416} {arXiv:hep-ph/9604416} \BibitemShut
  {NoStop}%
\bibitem [{\citenamefont {Oller}\ and\ \citenamefont
  {Oset}(1999)}]{Oller:1998zr}%
  \BibitemOpen
  \bibfield  {author} {\bibinfo {author} {\bibfnamefont {J.~A.}\ \bibnamefont
  {Oller}}\ and\ \bibinfo {author} {\bibfnamefont {E.}~\bibnamefont {Oset}},\
  }\Doi {10.1103/PhysRevD.60.074023} {\bibfield  {journal} {\bibinfo  {journal}
  {Phys. Rev. D},\ }\textbf {\bibinfo {volume} {60}},\ \bibinfo {pages}
  {074023} (\bibinfo {year} {1999})},\ \Eprint
  {http://arxiv.org/abs/hep-ph/9809337} {arXiv:hep-ph/9809337} \BibitemShut
  {NoStop}%
\bibitem [{\citenamefont {Giacosa}(2020)}]{Giacosa:2019ldb}%
  \BibitemOpen
  \bibfield  {author} {\bibinfo {author} {\bibfnamefont {F.}~\bibnamefont
  {Giacosa}},\ }\Doi {10.5506/APhysPolBSupp.13.83} {\bibfield  {journal}
  {\bibinfo  {journal} {Acta Phys. Polon. Supp.},\ }\textbf {\bibinfo {volume}
  {13}},\ \bibinfo {pages} {83} (\bibinfo {year} {2020})},\ \Eprint
  {http://arxiv.org/abs/1904.10368} {arXiv:1904.10368 [hep-ph]} \BibitemShut
  {NoStop}%
\bibitem [{\citenamefont {Lukashov}\ and\ \citenamefont
  {Simonov}(2020)}]{Lukashov:2019dir}%
  \BibitemOpen
  \bibfield  {author} {\bibinfo {author} {\bibfnamefont {M.~S.}\ \bibnamefont
  {Lukashov}}\ and\ \bibinfo {author} {\bibfnamefont {Y.~A.}\ \bibnamefont
  {Simonov}},\ }\Doi {10.1103/PhysRevD.101.094028} {\bibfield  {journal}
  {\bibinfo  {journal} {Phys. Rev. D},\ }\textbf {\bibinfo {volume} {101}},\
  \bibinfo {pages} {094028} (\bibinfo {year} {2020})},\ \Eprint
  {http://arxiv.org/abs/1909.10384} {arXiv:1909.10384 [hep-ph]} \BibitemShut
  {NoStop}%
\bibitem [{\citenamefont {Zheng}(2003)}]{Zheng:2003cr}%
  \BibitemOpen
  \bibfield  {author} {\bibinfo {author} {\bibfnamefont {H.-q.}\ \bibnamefont
  {Zheng}},\ }in\ \href@noop {} {\emph {\bibinfo {booktitle} {{International
  Symposium on Hadron Spectroscopy, Chiral Symmetry and Relativistic
  Description of Bound Systems}}}}\ (\bibinfo {year} {2003})\ pp.\ \bibinfo
  {pages} {98--105},\ \Eprint {http://arxiv.org/abs/hep-ph/0304173}
  {arXiv:hep-ph/0304173} \BibitemShut {NoStop}%
\end{thebibliography}%

\end{document}